\documentclass[particles,article,accept,moreauthors,pdftex]{Definitions/mdpi} 
\usepackage{amssymb,graphicx,bm,mathrsfs,color,amsmath,slashed,comment,isotope}
\newcommand{\bea}{\begin{eqnarray}}
\newcommand{\eea}{\end{eqnarray}}
\newcommand{\be}{\begin{equation}}
\newcommand{\ee}{\end{equation}}

\newcommand{\vecp}{{\bm p}}

\newcommand{\vecv}{{\bm v}}
\newcommand{\vecE}{{\bm E}}

\newcommand{\vecj}{{\bm j}}

\newcommand{\vecB}{{\bm B}}

\newcommand{\ep}{\varepsilon}

\newcommand{\ie}{{i.e.}}
\newcommand{\eg}{{e.g.}}

\input{./Definitions/acronym.input}
\definecolor{red}{rgb}{0.8,0,0}
\definecolor{violet}{rgb}{0.4,0,0.4}
\definecolor{green}{rgb}{0,0.5,0.0}
\definecolor{navy}{rgb}{0.0,0.0,0.6}
\definecolor{orange}{rgb}{0.8,0.2,0.0}
\usepackage[normalem]{ulem}  % \sout{old text} for strikeout                                                 

\firstpage{967} 
\pubvolume{7}
\issuenum{4}
\articlenumber{59}
\pubyear{2024}
\copyrightyear{2024}
\externaleditor{Academic Editor: Nicolas Chamel}
\datereceived{3 September 2024} 
\daterevised{3 November 2024} % Comment out if no revised date
\dateaccepted{7 November 2024} 
\datepublished{11 November 2024} 
\hreflink{https://doi.org/10.3390/\linebreak{}particles7040059} % If needed use \linebreak
\Title{Thermal Conductivity and Thermal Hall Effect in Dense Electron-Ion  Plasma}

\TitleCitation{Thermal Conductivity and Thermal Hall Effect in Dense Electron-Ion  Plasma}

\Author{{Arus Harutyunyan} $^{1,2,}$*$^{,\dagger,}$\orcidA{}  and Armen Sedrakian $^{3,4,\dagger}$\orcidB{}}
\AuthorNames{Arus Harutyunyan and Armen Sedrakian}

\AuthorCitation{{Harutyunyan}, A.; Sedrakian, A.} %\MDPI: Please check all author names carefully.

\address{%
$^{1}$ \quad Byurakan Astrophysical Observatory,  Byurakan 0213, Armenia\\%\MDPI: Please check that the address information is complete. The provided information should be arranged from subordinate to superior. Same below in aff 3.
$^{2}$ \quad Department of Physics, Yerevan State University, Yerevan 0025, Armenia\\
$^{3}$ \quad Frankfurt Institute for Advanced Studies, D-60438 Frankfurt am Main, Germany; {sedrakian@fias.uni-frankfurt.de} \\%\MDPI: We added the email addresses here according to those submitted online at susy.mdpi.com. Please confirm.
$^{4}$ \quad Institute of Theoretical Physics, University of Wroc\l{}aw, 50-204 Wroc\l{}aw, Poland}

\corres{Correspondence: arus@bao.sci.am}
\firstnote{These authors contributed equally to this work.} 

\abstract{In this study, we examine thermal conductivity and the  thermal Hall effect in electron-ion plasmas relevant to hot neutron  stars, white dwarfs, and binary neutron star mergers, focusing on  densities found in the outer crusts of neutron stars and the  interiors of white dwarfs.  We consider plasma consisting of single  species of ions, which could be either iron $\isotope[56]{Fe}$ or   carbon $\isotope[12]{C}$ nuclei.  The temperature range explored is from the  melting temperature of the solid $T\sim10^9$~K up to $10^{11}$~K.  This covers both degenerate and non-degenerate electron regimes. We  find that thermal conductivity increases with density and  temperature for which we provide analytical scaling relations valid in different regimes. The impact of magnetic fields on thermal  conductivity is also analyzed, showing anisotropy in low-density  regions and the presence of the thermal Hall effect characterized by the Righi--Leduc coefficient. The transition froma  degenerate to   non-degenerate regime is characterized by a minimum ratio  of thermal conductivity to temperature, which is analogous to the  minimum observed already in the case of electrical conductivity. We provide also formulas fit to our numerical results, which can be used in dissipative magneto-hydrodynamics simulations of warm \mbox{compact stars.}}

 %\MDPI: Please add the keywords.

\begin{document}

\section{Introduction}  

In the outer and inner crust of neutron stars and the cores of
white dwarfs, the plasma is composed primarily of degenerate electrons
and various ion species, depending on the density and
temperature. These ions might include light nuclei (like hydrogen
and helium) and heavier nuclei (iron), with possible other exotic nuclei deeper within the
crust of a neutron star.  In mature cold compact stars, the electrons
are highly degenerate, i.e., they fill a Fermi sphere. The transport
coefficients of such plasmas, including thermal conductivity, in the
liquid and solid phases, have been the focus of research for many
decades~\cite{Flowers1976,Yakovlev1980,Urpin1980,Flowers1981,Itoh1983,Itoh1984,Nandkumar1984MNRAS,Itoh1993,Baiko1998,Potekhin1999,Itoh2008,Potekhin2015}.
More recently, the transport in high-temperature, semi- or
non-degenerate regimes has become of great interest, mainly in the
context of binary neutron star mergers, but also proto-neutron
stars born in supernova explosions, which contain electron-ion plasma
that can be non-degenerate due to high temperatures. Understanding
transport in such an environment is needed for setting up dissipative
relativistic hydrodynamics, which ultimately determines the large-scale
behavior of these objects and allows us to model them numerically. Early
studies concentrated on electrical
conductivity~\cite{Harutyunyan2016,Harutyunyan2016b} and bulk
viscosity~\cite{Alford2019,Alford2020,Alford2021,Alford2023}. However,
the simultaneous presence of magnetic and electric fields and compositional
and thermal gradients couples the various transport channels. This
motivates our investigation of thermal conductivity effects within the same framework for electrical conductivity.

Neutron stars commonly have extremely strong magnetic fields, ranging
from canonical $10^{12}$~G to magnetar-strength $\ge$10\textsuperscript{15}~G
fields. Such strong fields drastically alter the behavior of charged
particles in their crusts. The magnetic field can cause anisotropies
in thermal and electrical conductivities as, for example, heat flows
more efficiently along magnetic field lines than across
them. Additionally, they lead to the Hall effect, where the motion of
electrons under the influence of a magnetic field generates transverse electric fields and/or temperature gradients. Furthermore, the
early stages of hot proto-neutron stars and post-merger remnants may
feature large thermal gradients driven by violent fluid dynamics.
The simultaneous presence of large magnetic fields and thermal gradients
can then lead to complex feedback loops between the temperature and
magnetic field evolution of such objects. Models were put forward
early on to describe the generation of strong fields via the dynamo effect
driven by strong thermal gradients, i.e., where the heat flux drives
the circulation that amplifies the magnetic field; this type of model is known as
thermoelectric~\cite{Dolginov1980,Blandford1983,Geppert1991,Wiebicke1996,Gakis2024}.
Another class of models concerns the thermo-magnetic evolution of
compact stars where their cooling is strongly affected by the presence
of strong magnetic fields, their decay, and internal
heating~\cite{Kaminker2006,Geppert2017,Dehman2023,Ascenzi2024}.  The third class of models describes neutron star binary mergers in the
presence of strong magnetic fields within resistive
magneto-hydrodynamics. Estimates of efficiency of the conductivity and
Hall effect in this context show the necessity of resistive
studies of mergers for proper understanding of the spectrum of
emitted gravitational waves~\cite{Harutyunyan2018}.

In this work, we compute the thermal conductivity tensor as well as
the Righi--Leduc coefficient describing the  thermal Hall effect in the
magnetized outer crust of a warm neutron star. In this domain of the
crust, the electrons are the main charge and heat carriers, and the
transport coefficients are dominated by the electron scattering off
the ions. The formalism we use here was previously developed for
assessment of electrical conductivity of electron-ion
plasma~\cite{Harutyunyan2016,Harutyunyan2016b,Harutyunyan2024}. Before
proceeding, let us point out that the thermal conductivity was
computed in the cold regime, including the case where ions solidify in the context of cold neutron stars in
Refs.~\cite{Flowers1976,Yakovlev1980,Flowers1981,Itoh1984,Sedrakian1987,Itoh1993,Shternin2006,Itoh2008} (see \cite{Potekhin2015} for a review). Work on the conductive opacity of low- and intermediate-mass stars has also been carried out ~\cite{Cassisi2007ApJ,Cassisi2021AA} within formalism applicable at arbitrary temperatures, densities, and magnetic fields. Their numerical work is  focused on lower densities ($\rho\le 10^8$ g\,cm$^3$), temperatures ($T\le 10^8$~K), and light element (H and He) composition.

This paper is organized as follows. Section~\ref{sec:regimes} discusses the phase diagram of one-component plasma in the regimes of interest for neutron stars and white dwarfs. In Section~\ref{sec:Boltzmann}, we present the computation energy-dependent relaxation time starting from the Boltzmann equation and the key approximations that underlie our formalism. In Section~\ref{sec:currents},
we derive the tensors of the thermal conductivity in magnetic fields by evaluating the electric and thermal current when the  electric field and temperature gradient are present  simultaneously.
Section~\ref{sec:results} presents the numerical results for the thermal conductivity and the thermal Hall effect in the density, temperature, and $B$-field regimes of interest. Our results are summarized in Section~\ref{sec:conclusions}. 

We use the natural (Gaussian) units with $\hbar= c = k_B = k_e = 1$, $e=\sqrt{\alpha}$, and 
$\alpha=1/137$, and the metric signature
$(1,-1,-1,-1)$.

\section{Physical Conditions}
\label{sec:regimes}

\begin{figure}[H] 
%\begin{center}
\includegraphics[width=0.5\linewidth,keepaspectratio]{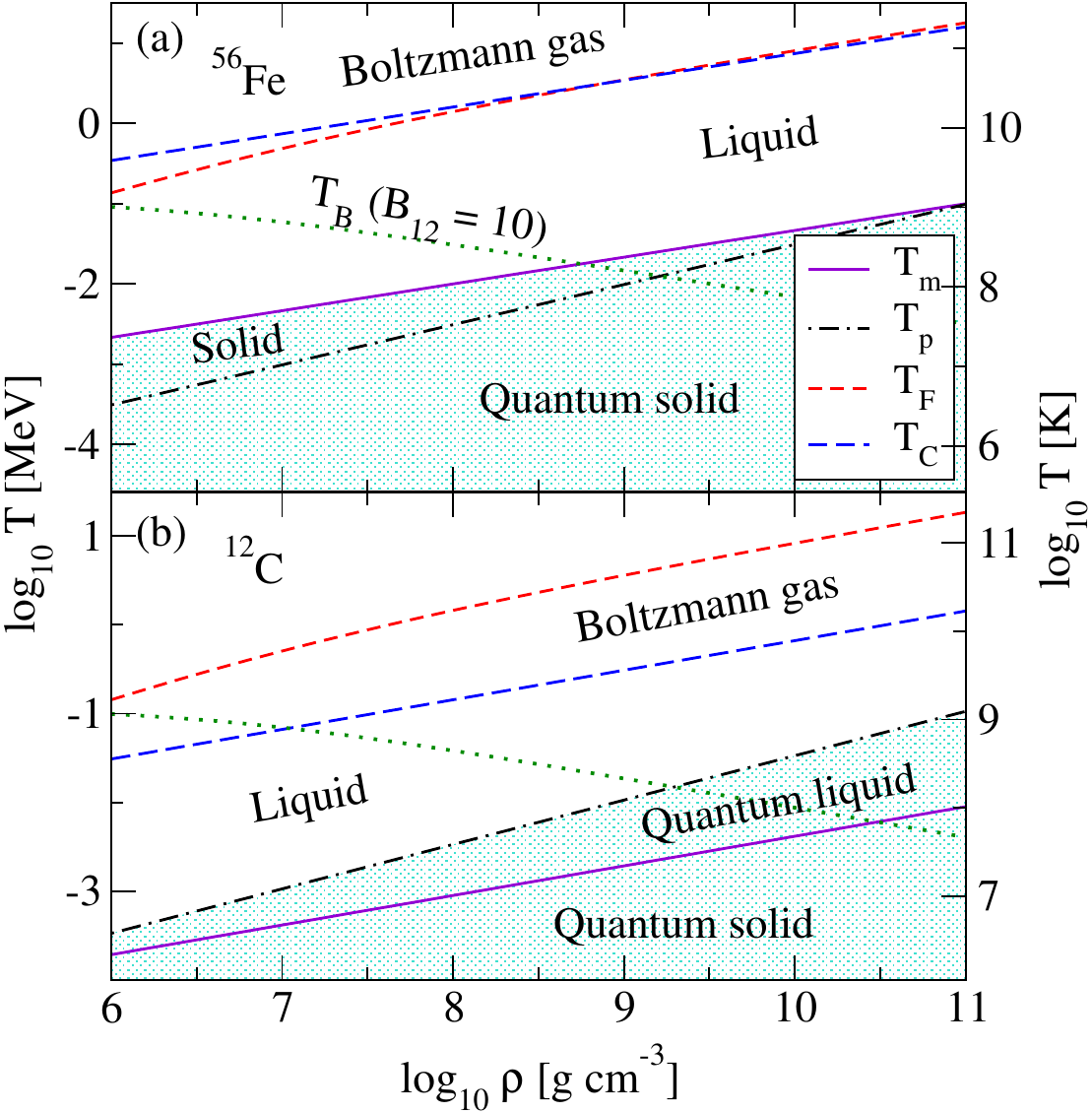}
\caption{ The temperature-density phase diagram of dense plasma
  composed of iron $\isotope[56]{Fe}$ (\textbf{a}) and carbon
  $\isotope[12]{C}$ (\textbf{b}). The electron gas degeneracy sets in
  below the Fermi temperature $T_F$ (short dashed lines). The ionic
  component solidifies below the melting temperature $T_m$ (solid
  lines), while quantum effects become important below the plasma
  temperature (dash-dotted lines).  For temperatures above $T_{\rm C}$
  (long dashed lines), the ionic component forms a Boltzmann gas. 
  Note
  that for $\isotope[12]{C}$, quantum effects become important
  in the portion of the phase diagram lying between the lines
  $T_p(\rho)$ and $T_m(\rho)$. 
In addition, the dotted lines show the temperature $T_B$ below which the quantization of the electron energy in the magnetic field becomes important for $B_{12}=10$.
  {The} present study does not cover the shaded portion of the phase diagram. }%\MDPI: We set this sentence as figure caption, please confirm. Please change the hyphen (-) into a minus sign (−, “U+2212”) in the figure, e.g., “-1” should be “−1”. We moved this figure after it first citation in the whole text, please check and confirm
\label{fig:PhaseDiagram} 
%\end{center}
\end{figure}

Matter in the outer crusts of neutron stars and the interiors of 
white dwarfs consists of fully ionized ions (\ie, atomic nuclei) and 
almost non-interacting Fermi gas of relativistic electrons, which are 
the main transporters of charge and heat in the system. Electron ($n_e$)
and ion ($n_i) $densities are related by the charge local neutrality condition
$n_e=Zn_i$, where $Z$ is the ion charge number. The almost free gas of electrons 
becomes degenerate below the Fermi temperature $T_F = \varepsilon_F-m$, 
where $\varepsilon_F=(p_F^2+m^2)^{1/2}$ is the Fermi energy, $p_F = 
(3\pi^2n_e)^{1/3}$ is the Fermi momentum, and $m$ is the electron mass. 
The state of the ionic component is characterized by the so-called Coulomb 
plasma parameter %with mass number $A$ 
\vspace{-3pt}
%------------------------------------------------------
\bea\label{eq:Gamma}
\Gamma=\frac{T_C}{T},\quad T_C=\frac{e^2 Z^2}{a_i},
\eea
%------------------------------------------------------
where $e$ is the elementary charge, $T$ is the temperature, and
$a_i=(4\pi n_i/3)^{-1/3}$ is the radius of the spherical volume 
per ion. If $\Gamma\ll 1$, or, equivalently, $T\gg T_{\rm C}$, ions form weakly 
coupled Boltzmann gas. In the opposite regime, $\Gamma\ge 1$ ions 
are strongly coupled and form either a liquid (for 
$\Gamma\leq\Gamma_m\simeq 160$) or a lattice (for $\Gamma>\Gamma_m$). 
Correspondingly, the melting temperature of the lattice is defined 
as $T_m=(Ze)^2/\Gamma_ma_i$. The ion plasma temperature is represented by\vspace{-3pt}
%------------------------------------------------------------------
\bea
  T_p = \biggl(\frac{4\pi  Z^2e^2n_i}{M }\biggl)^{1/2},
\eea
%------------------------------------------------------------------
where $M $ is the ion mass, and further separates the regimes of classical
($T\geq T_p$) and quantum ($T\leq T_p$) lattice.  In the second case,
the quantization of the lattice oscillations becomes important. Figure
\ref{fig:PhaseDiagram} shows the temperature-density phase diagram of
the crustal plasma composed of iron $\isotope[56]{Fe}$ and carbon $\isotope[12]{C}$ nuclei.
%----------------------------------------------------------------

%\vspace{-3pt}
%----------------------------------------------------------------

The general structure of the phase diagram is similar for
$\isotope[56]{Fe}$ and carbon $\isotope[12]{C}$ nulcei. However, an important difference appears in the low-temperature regime: while the
plasma temperature is always smaller than the melting temperature in
the case of $\isotope[56]{Fe}$, in the case of $\isotope[12]{C}$
instead, the quantum effects become important before the solidification
sets in. In~addition, the phase diagrams for these elements do not take into account that at high temperatures and/or densities, the energy and pressure are sufficient to overcome the Coulomb barrier, allowing nuclei to fuse almost instantaneously. This fusion process occurs more rapidly than the rates at which the plasma can maintain its low-$Z$ composition. As a result, rather than remaining in a stable plasma state, the particles in the plasma will quickly undergo fusion reactions, leading to the formation of heavier elements. Consequently, the existence of a low-$Z$ plasma becomes untenable in such environments.

We note also that the phase diagram shown in Figure~\ref{fig:PhaseDiagram} is applicable to the non-magnetized crust. In the presence of magnetic fields, the Fermi temperature of electrons is modified at low densities where the Landau quantization is important. This is the case at temperatures $T\leq T_B$, with\vspace{-3pt}
  %------------------------------------------------------------------
\bea
  T_B = \frac{eB}{\varepsilon_F},
\eea
%------------------------------------------------------------------
  which is the characteristic temperature which separates the 
  quantized and non-quantized regimes of the phase diagram~\cite{Potekhin1999}. 
  The temperature $T_B$ is shown in Figure~\ref{fig:PhaseDiagram}, with dotted lines for the magnetic field  $B_{12}=10$. As seen from the figure, for $B_{12}\leq 10$, the Landau quantization can be neglected at temperatures $T\geq 0.1$~MeV, which is the main regime of our interest. However, for higher magnetic fields, the quantization of electron energies becomes important in the low density domain of the phase diagram. In this work, we will restrict our analysis to the non-quantizing regime, keeping in mind that the extrapolation of the obtained results to higher values of magnetic field should be regarded as the classical average values of the transport coefficients in the regime where the quantization \mbox{is important.}

\section{Thermal Conductivity and Thermal Hall Effect from Boltzmann Equation}
\label{sec:Boltzmann}

We start with the Boltzmann equation 
for the electron distribution {function} %\MDPI: Please check through the full text that if all bolds in equations are necessary.
\vspace{-3pt}
%-----------------------------------------------------------
\bea\label{eq:boltzmann}
\frac{\partial f}{\partial t}+
\bm v\frac{\partial f}{\partial\bm r}-
e(\bm E+\bm v\times \bm B)\frac{\partial f}
{\partial\bm p}=I[f],
\eea
%-----------------------------------------------------------
where $\vecE$ and $\vecB$ are the electric and magnetic fields, 
$\bm v =\partial\ep/\partial \bm p$ is the electron velocity with
$\ep=\sqrt{p^2+m^2}$ and 
$m$ being the electron mass, $e$ is the unit charge, and $I[f]$ is 
the electron-ion collision integral given by
\vspace{-3pt}
%-----------------------------------------------------------
\bea\label{eq:collision_ei}
I[f]=-(2\pi)^4\sum\limits_{234}|{\cal M}_{12\to 34}|^2 
\delta^{(4)}(p_1+p_2-p_3-p_4)
[f_1(1-f_3)g_2-f_3(1-f_1)g_4],
\eea
%-----------------------------------------------------------
where $f_{1,3}=f(p_{1,3})$ are the electrons' and $g_{2,4}=g(p_{2,4})$ are the 
ions' distribution functions, 
${\cal M}_{12\to 34}$
is the electron-ion scattering matrix element,
and $\sum\limits_{i}=\int d\bm p_i/(2\pi)^3$.
We will refer the study of thermal conductivity resulting from electron--electron collisions to a future work, as its contribution is minor to the nuclear charge numbers  considered here, see also~\cite{Shternin2006,Cassisi2007ApJ,Cassisi2021AA}.

As ions form a classical fluid in equilibrium,  the function $g(p)$ is given by the  Maxwell--Boltzmann distribution\vspace{-3pt}
%------------------------------------------------------
\bea\label{eq:maxwell}
g(p)=n_i\bigg(\frac{2\pi}{MT}\bigg)^{3/2}
\exp\left(-\frac{p^2}{2MT}\right).
\eea
%------------------------------------------------------

{To} linearize the kinetic equation, the electron %\MDPI: We revised the indentation format, please confirm. Please check all similar revisions in the full text.
 distribution function can be written in the form\vspace{-3pt}
%-----------------------------------------------------------
\bea\label{eq:distribution}
f= f^0+\delta f,\quad \delta f=-\phi
\frac{\partial f^0}{\partial\ep}.
\eea
%-----------------------------------------------------------
where\vspace{-3pt}
%-----------------------------------------------------------
\bea\label{eq:fermi}
f^0({\ep})=\frac{1}{e^{\beta(\ep -\mu)}+1}
\eea
%-----------------------------------------------------------
is the local Fermi distribution function with $\beta=T^{-1}$, $\mu$ is the electron chemical potential,  and $\delta f\ll f^0$ is a small perturbation.  To derive the thermal 
conductivity, we assume that the temperature $T$ and the electron 
chemical potential $\mu$ are slowly varying functions of space:
$T=T(\bm x)$, $\mu=\mu(\bm x)$.  Using Equation~\eqref{eq:fermi}, the space
and momentum 
derivatives can be computed in the drift term of Equation~\eqref{eq:boltzmann}.
% -----------------------------------------------------------
Note that in Equation~\eqref{eq:boltzmann}, the substitution
$f \to f^0$ can be applied to the terms involving space-time
derivatives or the electric field. However, in the term involving the
magnetic field, the perturbation must be retained because, in
equilibrium, the magnetic field's effect vanishes since $[\vecv\times \vecB]({\partial f^0}/{\partial\vecp})\propto[\vecv\times \vecB]\vecv=0.$
Thus, for the third term in Equation~\eqref{eq:boltzmann},
we obtain\vspace{-3pt}
%-----------------------------------------------------------
\bea\label{eq:field_term}
e(\bm E+\bm v\times \vecB)\frac{\partial f}
{\partial\bm p}=e\bm v\cdot\bm E\frac{\partial f^0}
{\partial\ep}-e[\bm v\times\vecB]
\frac{\partial f^0}{\partial\ep}
\frac{\partial\phi}{\partial\bm p}.
\eea
%-----------------------------------------------------------

{Using} the space- and momentum-derivatives of Equations~\eqref{eq:fermi} 
and \eqref{eq:field_term}, we obtain the left-hand-side of
Equation~\eqref{eq:boltzmann}
at linear order in macroscopic gradients \vspace{-3pt}
%------------------------------------------------------------
\bea\label{eq:boltzmann_linear_1}
\frac{df}{dt}
= -\frac{\partial f^0}{\partial \ep}
\left(e\bm v\cdot\bm F
-e[\bm v\times\vecB]\frac{\partial\phi}{\partial\bm p}\right),
\eea
%-----------------------------------------------------------
where we define\vspace{-3pt}
%--------------------------------------------------------------
\bea\label{eq:vector_F}
\bm F=\bm E'+\frac{\ep-\mu}{eT}\nabla T,
\qquad e\bm E'=e\bm E+\nabla \mu.
\eea
%--------------------------------------------------------------

{Thus}, the linearized Boltzmann equation reads\vspace{-3pt}
%--------------------------------------------------------------
\bea\label{eq:boltzmann_linear}
\frac{\partial f^0}{\partial \ep}
\left(e\bm v\cdot\bm F
-e[\bm v\times\vecB]\frac{\partial\phi}{\partial\bm p}\right)=-I[\phi],
\eea
%--------------------------------------------------------------
with the linearized collision integral given by\vspace{-3pt}
%-----------------------------------------------------------
\bea\label{eq:coll_ei_lin}
I[\phi] = -(2\pi)^4\beta\sum\limits_{234}
|{\cal M}_{12\to 34}|^2
\delta^{(4)}(p_1+p_2-p_3-p_4)
f^0_1(1-f^0_3)g_2(\phi_1-\phi_3).
\eea
%-----------------------------------------------------------

{The} solution of Equation~\eqref{eq:boltzmann_linear_1} can be obtained from
Equation~(15) of Ref.~\cite{Harutyunyan2016} by substitution
$\bm E\to \bm F$\vspace{-3pt}
%-----------------------------------------------------------
\bea\label{eq:psi_cond}
\phi = -\frac{e\tau}{1+(\omega_c\tau)^2}
v_i\left(\delta_{ij}-\omega_c\tau\varepsilon_{ijk}
b_k+(\omega_c\tau)^2b_ib_j\right)F_j,
\eea
%-----------------------------------------------------------
where $\bm b=\vecB/B$, $\omega_c=eB/\ep$ is the electron cyclotron frequency, and  the relaxation time is defined by \vspace{-3pt}
%---------------------------------------------------------------
\bea\label{eq:t_relax}
\tau^{-1}(\ep_1)=(2\pi)^{-5}\!
\int\! d\bm q\!\int\! d\bm p_2\,
|{\cal M}_{12\to 34}|^2 
\frac{\bm q\cdot \bm p_1}{p^2_1}
\delta(\ep_1+\ep_2-\ep_3-\ep_4)
g_2\frac{1-f^0_3}{1-f^0_1}.
\eea
%----------------------------------------------------------------------

{The} matrix element includes several corrections to the bare Coulomb electron-ion interaction.  The screening of the interaction is included in terms of the hard-thermal-loop polarization tensor of QED
plasma in the low-frequency limit (see Sec.~IV D of
Ref.~\cite{Harutyunyan2016}). 
The ion--ion correlations are accounted for by means of the static structure factor $S(q)$ of one-component plasma,
and the finite nuclear size is included via the nuclear form factor
$F(q)$ (see Ref.~\cite{Harutyunyan2016} and the discussion below). The
final expression for the relaxation \mbox{time reads}\vspace{-3pt}
%------------------------------------------------------------------
\bea\label{eq:relax_final}
\tau^{-1}(\ep) & =&\frac{\pi Z^2 e^4 n_i}{\ep p^3} \sqrt{\frac{M}{{2 \pi}T }} \!\int_{-\infty}^{\ep-m}\! d \omega\, e^{-\omega / 2 T} \frac{f^0(\ep-\omega)}{f^0(\ep)} \int_{q_{-}}^{q_{+}}\! d q\left(q^2-\omega^2+2 \ep \omega\right) \nonumber\\
 &\times &
 e^{-\omega^2 M/ 2 q^2 T} e^{-q^2 / 8 M T} S(q) F^2(q)
%\bigg[
\frac{(2 \ep-\omega)^2-q^2}{\left|q^2+\Pi_L\right|^2},
%+ \frac{T}{M}\frac{(q^2-\omega^2)[(2\ep-\omega)^2
%+q^2]-4m^2q^2}{q^2|q^2-\omega^2+\Pi_T|^2}\bigg],
\eea
%------------------------------------------------------------------
where $q_{\pm} =\vert \sqrt{p^2-(2\omega\ep- \omega^2)} \pm p\vert$, and the real and imaginary parts of the longitudinal 
polarization are given by
$\Pi_L (q,\omega) = q_D^2\chi_l$,
%$\Pi_T (q,\omega) = q_D^2\chi_t$ 
where\vspace{-3pt}
%---------------------------------------------------------
\bea\label{eq:chi_l}
&&{\rm Re}\chi_l (q,\omega) 
\simeq 1-\left(\frac{\omega}{q}\right)^2, %\frac{x^2}{\bar{v}^2},
\quad {\rm Im}\chi_l (q,\omega) 
\simeq -\frac{\pi \omega}{2q}, 
%\frac{\pi x}{2\bar{v}},
%\label{eq:chi_t}
%&&{\rm Re}\chi_t \simeq x^2,  \qquad 
%{\rm Im}\chi_t \simeq \frac{\pi}{4}x,
\eea 
%---------------------------------------------------------
where the 
 Debye wave-number $q_D$ at finite temperature is given by \vspace{-3pt}
%-------------------------------------------------------------
\bea\label{eq:Debye}
q_D^{2} =-\frac{4e^2}{\pi} \int_0^{\infty}\!\! 
dp\,p^2\frac{\partial f^0}{\partial\varepsilon}.
\eea
%-------------------------------------------------------------

{Note} that we neglected the transverse part of the scattering in the matrix element as its contribution to the relaxation time is small in the regime of interest of this work (for the full expression, see \cite{Harutyunyan2016}).  
{Equations~\eqref{eq:relax_final}--\eqref{eq:Debye} are valid for arbitrary temperatures and across the full spectrum  of degeneracies of electrons. 

The relaxation time \eqref{eq:relax_final} takes a simpler form when the ionic component of the plasma is considered at zero temperature and nuclear recoil is neglected  (see Appendix A of \cite{Harutyunyan2016})\vspace{-3pt}:
%-------------------------------------
\bea\label{eq:Coulomb_log}
\tau^{-1}(\varepsilon)=\frac{4\pi Z^2 e^4 n_i}{\varepsilon p^3} 
\int_0^{2 p} d q q^3 S(q) F^2(q) \frac{ \varepsilon^2-(q/2)^2}{\left|q^2+\Pi_L(0,q)\right|^2}.
\eea
%-------------------------------------

{The} integral in this expression is known as {\it {C}oulomb logarithm (CL)}~\cite{Lee1950,Hubbard1969,Yakovlev1980}. It has been widely used in the previous literature to accommodate the influence of correlations on the transport in electron-ion plasma. In this form, it  can be applied at arbitrary temperatures, as long as the finite-temperature effects on the ionic component and recoil can be neglected (\mbox{see \cite{Potekhin1999,Cassisi2007ApJ,Cassisi2021AA}}).
The many-body approximation enters the CL via the ion structure factor $S(q, \omega)$, which describes the correlations in the ionic component and the longitudinal polarization tensor of the electrons $\Pi_L(\omega,q)$. The state of the art computations 
of $S(q,\omega)$, typically taken in the static limit $\omega=0$, are based on hypernetted-chain or Monte Carlo methods for classical one-component plasma. Itoh et al.~\cite{Itoh1983} implemented in their low-temperature thermal conductivity evaluation the results of hypernetted-chain computations of static structure factor for various values of the $\Gamma$ parameter, and compared their results for the CL to that of Ref.~\cite{Yakovlev1980} for $^{56}$Fe and $^{12}$C. They also implemented the low-temperature random-phase-approximation polarization tensor for relativistic electrons in the static ($\omega=0$) limit given by ~\cite{Jancovici1962}. The finite-temperature computations of \cite{Cassisi2007ApJ,Cassisi2021AA}  for low-mass stars used a fit formula for CL derived in \cite{Potekhin1999}, which is informed by fits to  hypernetted chain  computations of one-component plasma given in \cite{Young1991}. This fit formula also includes screening through its dependence on the Thomas--Fermi and Debye wave-lengths. } {In our computations, we adopted the same approach as in \cite{Harutyunyan2016}. Specifically, 
the static structure factor is taken from the fits to Monte Carlo simulations of a one-component plasma~\cite{Hansen1973,Galam1976,Tamashiro1999}. The  
screening is accounted through hard-thermal-loop QED polarization 
functions in the low-frequency limit. This formalism does not 
impose any restrictions on the temperature or degeneracy of 
the electrons. }

\section{Electric and Thermal Currents}
\label{sec:currents}

We proceed now to the computation of the
electrical and thermal currents using the expression
\eqref{eq:vector_F} for $\bm F$. Using the standard kinetic theory definitions we find~\cite{Ziman1979} 
\vspace{-3pt}
%-----------------------------------------------------------
\bea
\label{eq:current_el}
j_k &=& \int\frac{2d\bm p}{(2\pi)^3}ev_k\phi
\frac{\partial f^0}{\partial\ep}=\sigma_{kj}E'_j-\alpha_{kj}
\partial_j T,
\eea
and \vspace{-3pt}
\bea
\label{eq:current_therm}
q_k &=& -\int\frac{2d\bm p}{(2\pi)^3}(\ep-\mu)v_k\phi
\frac{\partial f^0}{\partial\ep}=\tilde{\alpha}_{kj}E'_j
-\tilde{\kappa}_{kj}\partial_j T,
\eea
where\vspace{-3pt}
\bea
\sigma_{kj}&=& -\int\frac{2d\bm p}{(2\pi)^3}
\frac{\partial f^0}{\partial\ep}
\frac{e^2\tau}{1+(\omega_c\tau)^2}
v_kv_i\left[\delta_{ij}-\omega_c\tau\ep_{ijm}
b_m +(\omega_c\tau)^2b_ib_j\right]
\\
\alpha_{kj}&=& -\int\frac{2d\bm p}{(2\pi)^3}
\frac{\partial f^0}{\partial\ep}
\frac{e(\ep-\mu)\tau}{1+(\omega_c\tau)^2}
v_kv_i\left[\delta_{ij}-\omega_c\tau\ep_{ijm}
b_m +(\omega_c\tau)^2b_ib_j\right]T^{-1}\\
\tilde{\alpha}_{kj}
&=&\int\frac{2d\bm p}{(2\pi)^3}
\frac{\partial f^0}{\partial\ep}
\frac{e(\ep-\mu)\tau}{1+(\omega_c\tau)^2}
v_kv_i\left[\delta_{ij}-\omega_c\tau\ep_{ijm}
b_m +(\omega_c\tau)^2b_ib_j\right]
\\
\tilde{\kappa}_{kj}&=&-\int\frac{2d\bm p}{(2\pi)^3}
\frac{\partial f^0}{\partial\ep}
\frac{(\ep-\mu)^2\tau}{1+(\omega_c\tau)^2}
v_kv_i\left[\delta_{ij}-\omega_c\tau\ep_{ijm}
b_m +(\omega_c\tau)^2b_ib_j\right]T^{-1}.
\eea
%-----------------------------------------------------------

{{These} expressions for transport coefficients are valid for arbitrary temperatures (\mbox{see \cite{Ziman1979}}). They have been employed earlier in the context of low-temperature neutron stars and in the cases of non-quantizing~\cite{Urpin1980} and quantizing~\cite{Hernquist1984,Potekhin1999,Potekhin2015} magnetic fields and finite-temperature low-mass stars~\cite{Cassisi2007ApJ,Cassisi2021AA}. Note the simplification in the low-temperature limit, ${\partial f^0}/{\partial \varepsilon} \rightarrow -\delta\left(\varepsilon-\varepsilon_F\right)$, which removes the necessity for numerical integration.}

We will next assume without loss of generality that the magnetic field is directed along the $z$ axis. Inverting Equations~\eqref{eq:current_el} and \eqref{eq:current_therm} we
 find \vspace{-3pt}
%-----------------------------------------------------------
\bea\label{eq:currents_reversed}
\bm E' = \hat{\rho}\bm j
-\hat{Q}\nabla T,\qquad
\bm q = -\hat{\kappa}\nabla T
-T\hat{Q}\bm j,
\eea
%-----------------------------------------------------------
where $\hat{\rho}=\hat{\sigma}^{-1}$ and $\hat{\sigma}$ are the
matrices of electrical resistivity and conductivity, respectively;
$\hat{Q}=-\hat{\rho}\hat{\alpha} $ is the matrix of thermopower; and
$\hat{\kappa} =\hat{\tilde{\kappa}}+ T\hat{\alpha} \hat{Q} $ is the
matrix of thermal conductivity. These matrices of the transport
coefficients are given explicitly by 
%-----------------------------------------------------------
\vspace{-9pt}
\begin{adjustwidth}{-\extralength}{0cm}
%\centering %% If there is a figure in wide page, please release command \centering
  \bea\label{eq:cond_matrices}
&&\hat{\sigma} =
\begin{pmatrix}
   \sigma_0 & -\sigma_1 & 0 \\
    \sigma_1 & \sigma_0 & 0 \\
    0 & 0 & \sigma
\end{pmatrix},
\quad
\hat{\alpha} =
\begin{pmatrix}
    \alpha_0 & -\alpha_1 & 0 \\
    \alpha_1 & \alpha_0 & 0 \\
    0 & 0 & \alpha
  \end{pmatrix},  \nonumber\\
&& \hat{\tilde{\kappa}} =
\begin{pmatrix}
    \tilde{\kappa}_0 & -\tilde{\kappa}_1 & 0 \\
    \tilde{\kappa}_1 & \tilde{\kappa}_0 & 0 \\
    0 & 0 & \tilde{\kappa}
  \end{pmatrix}, 
  \quad
 \hat{\tilde{Q}}  =\begin{pmatrix}
    Q_0 & -Q_1 & 0 \\
    Q_1 & Q_0 & 0 \\
    0 & 0 & Q
\end{pmatrix},
\eea
\end{adjustwidth}
%-----------------------------------------------------------
with components given by \vspace{-3pt}
%-----------------------------------------------------------
\bea\label{eq:sigma_scalar}
\sigma &=& -\frac{e^2}{3\pi^2}
\int_m^\infty\! d\ep\,\frac{p^3}{\ep}
\frac{\partial f^0}{\partial\ep}\tau
=\sigma_0+\sigma_2,\\
\label{eq:sigmas}
\sigma_l &=& -\frac{e^2}{3\pi^2}
\int_m^\infty\! d\ep\,\frac{p^3}{\ep}
\frac{\partial f^0}{\partial\ep}\frac{\tau(\omega_c\tau)^l}
{1+(\omega_c\tau)^2},\\
\label{eq:alpha_scalar}
\alpha &=& \frac{e}{3\pi^2}
\int_m^\infty\! d\ep\,\frac{p^3}{\ep}\frac{\partial f^0}{\partial\ep}
\left(\frac{\ep-\mu}{T}\right)\tau
=\alpha_0+\alpha_2,\\
\label{eq:alphas}
\alpha_l &=& \frac{e}{3\pi^2}\int_m^\infty\! d\ep\,
\frac{p^3}{\ep}\frac{\partial f^0}{\partial\ep}
\left(\frac{\ep-\mu}{T}\right)\frac{\tau(\omega_c\tau)^l}
{1+(\omega_c\tau)^2},\\
\label{eq:til_kappa_scalar}
\tilde{\kappa} &=& -\frac{T}{3\pi^2}
\int_m^\infty\! d\ep\,\frac{p^3}{\ep}
\frac{\partial f^0}{\partial\ep}
\left(\frac{\ep-\mu}{T}\right)^2\tau
=\tilde{\kappa}_0+\tilde{\kappa}_2,\\
\label{eq:til_kappas}
\tilde{\kappa}_l &=& -\frac{T}{3\pi^2}
\int_m^\infty\! d\ep\,\frac{p^3}{\ep}
\frac{\partial f^0}{\partial\ep}
\left(\frac{\ep-\mu}{T}\right)^2
\frac{\tau(\omega_c\tau)^l}
{1+(\omega_c\tau)^2},
\eea
% -----------------------------------------------------------
which allows us to write down the components of the thermopower\vspace{-3pt}
% -----------------------------------------------------------
\bea\label{eq:Q_comp}
Q =-\frac{\alpha}{\sigma},\qquad
Q_0 =-\frac{\alpha_0\sigma_0 + \alpha_1\sigma_1}{\sigma_0^2+\sigma_1^2},\qquad
Q_1 = -\frac{\alpha_1\sigma_0 -\alpha_0\sigma_1}{\sigma_0^2+\sigma_1^2}.
\eea
% -----------------------------------------------------------

{Our} numerical computations below concentrate on the thermal conductivity given by the matrix \vspace{-3pt}
%-----------------------------------------------------------
\bea\label{eq:kappa_matrix}
\hat{\kappa}  =
\begin{pmatrix}
    \kappa_0 & -\kappa_1 & 0 \\
    \kappa_1 & \kappa_0 & 0 \\
    0 & 0 & \kappa
\end{pmatrix},
\eea
%-----------------------------------------------------------
where the coefficients are defined as \vspace{-3pt}
%-----------------------------------------------------------
\bea\label{eq:kappa_scalar}
\kappa &=& \tilde{\kappa}+T\alpha Q
=\tilde{\kappa}-T\frac{\alpha^2}{\sigma},\\
\label{eq:kappa_0}
\kappa_0 &=& \tilde{\kappa}_0+
T(\alpha_0 Q_0 - \alpha_1 Q_1),\\
\label{eq:kappa_1}
\kappa_1 &=& \tilde{\kappa}_1
+T(\alpha_0 Q_1 +\alpha_1 Q_0),
\eea
%-----------------------------------------------------------
where $\kappa$ is the 
longitudinal,
$\kappa_0$ is the transverse, and $\kappa_1$ is the Hall
component of the thermal conductivity. 
Note that the components of the thermal conductivity tensor are fully
determined if the relaxation time $\tau$ is known. We evaluate the
scattering matrix of electrons by ions using the standard QFT methods.
The many-body correlations are taken into account through the
structure-function of ions determined from Monte Carlo simulations of
one-component plasma and screening of ionic charge in the hard-thermal
loop approximation given by Equation~\eqref{eq:chi_l}. We also take into account the finite size of ions
by multiplying the matrix element with a suitable function of the radius
of the nucleus (see Ref.~\cite{Harutyunyan2016, Harutyunyan2024} for
details).  Finally, we cast Equation~\eqref{eq:currents_reversed} in the
following form\vspace{-3pt}:
%-----------------------------------------------------------
\bea\label{eq:E_vector_form}
\bm E' 
&=& {\varrho}\bm j_{\parallel}
+{\varrho}_0\bm j_{\perp}
+R[\bm B\times \vecj]
-{Q}\nabla_{\parallel} T
-{Q}_0\nabla_{\perp} T
+N[\bm B\times\nabla T],\\
\label{eq:q_vector_form}
\bm q 
&=& -{\kappa}\nabla_{\parallel} T
-{\kappa}_0\nabla_{\perp} T
+L[\bm B\times\nabla T]
-T{Q}\bm j_{\parallel}-T{Q}_0\bm j_{\perp}
+D[\bm B\times\bm j],
\eea
%-----------------------------------------------------------
where $\parallel$ and $\perp$ denote the components of vectors that
are parallel and perpendicular to the magnetic field with\vspace{-3pt}
%-----------------------------------------------------------
\bea\label{eq:rho_comp}
\varrho =\frac{1}{\sigma},\qquad
\varrho_0 =\frac{\sigma_0}{\sigma_0^2+\sigma_1^2},\qquad
\varrho_1 = \frac{\sigma_1}{\sigma_0^2+\sigma_1^2},
\eea
%-----------------------------------------------------------
denoting the components of the electrical resistivity tensor
$\hat{\varrho}={\hat{\sigma}}^{-1}$.

\textls[-15]{Here, we introduced
the Hall coefficient $R=-\rho_1/B$, the Nernst coefficient \mbox{$N=-Q_1/B$},}
the Righi--Leduc coefficient $L=-\kappa_1/B$, and the Ettingshausen
coefficient $D=-TQ_1/B=NT$. The physical content of each term is
easily read off from the corresponding terms in the currents
\eqref{eq:E_vector_form} and \eqref{eq:q_vector_form}. Below, we focus on the thermal conductivity and the thermal Hall effect, which describes the heat flow
orthogonal to the temperature gradient under an imposed magnetic
field. Its efficiency is measured by the Righi--Leduc coefficient
$L$. Clearly, it is an analog of the electrical Hall effect ($\propto R$
in Equation~\eqref{eq:E_vector_form}),
which occurs when a magnetic field is applied perpendicular to an
electric current in a conductor, causing the charge carriers
(electrons) to create a voltage across the system.

\section{Results}
\label{sec:results}

In this section, we focus on our results for thermal conductivity and
thermal Hall effect. Numerically, these are evaluated using the
relaxation time Equation~\eqref{eq:relax_final} and the formulas in
Equations~\eqref{eq:sigma_scalar}--\eqref{eq:Q_comp}, and
\eqref{eq:kappa_scalar}--\eqref{eq:kappa_1}.  We recall that in large
magnetic fields ($\omega_c\tau \gtrsim 1$) the tensor structure of
these quantities becomes important, while for low magnetic fields
($\omega_c\tau \to 0$) only the longitudinal thermal conductivity
$\kappa$ is relevant.  In the isotropic case $\omega_c\tau \ll 1$, one
has $\sigma_1\ll \sigma_0\simeq\sigma$; therefore, all three diagonal
components of the conductivity tensor are identical, and the
non-diagonal components vanish. In the anisotropic case
$\omega_c\tau \simeq 1$, they are distinct and should be studied
separately.  Below, we will study the dependence of the conductivity
on the density, temperature, and strength of the magnetic field for
the selected compositions.

\subsection{Results for Relaxation Times}
\label{app:tau}

\begin{figure}[H] 
%\begin{center}
\includegraphics[width=0.75\linewidth,keepaspectratio]{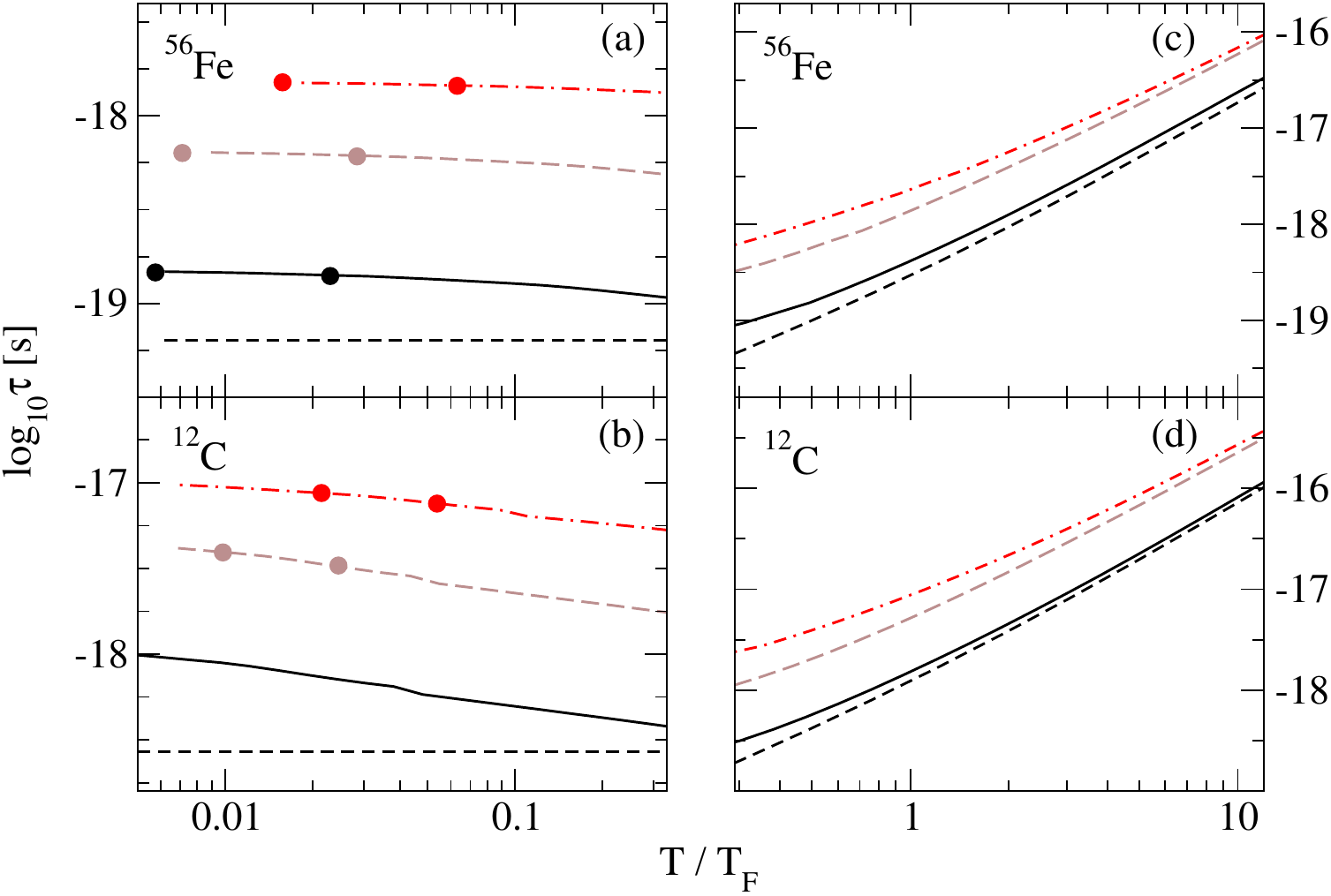}
\caption{{The} relaxation time as a function of dimensionless ratio
  $T/T_F$ for three values of density: $\log_{10}\rho=10$ (solid
  lines), $\log_{10}\rho=8$ (dashed lines), and $\log_{10}\rho=6$
  (dash-dotted lines) for $\isotope[56]{Fe}$ [(\textbf{a},\textbf{c})] and
  $\isotope[12]{C}$ [(\textbf{b},\textbf{d})]. Panels (\textbf{a},\textbf{b})
  correspond to the degenerate regime, and  (\textbf{c},\textbf{d}) to the
  nondegenerate regime.   
  To demonstrate the effect of ion--ion correlations, we show the results for $\log_{10}\rho=10$ in the case where $S(q) = 1$ by short-dashed lines.
    The circles reproduce the results of Ref.~\cite{Nandkumar1984MNRAS}.}
\label{fig:tau_temp1}%\MDPI: We moved this figure after it first citation in the whole text, please check and confirm.  Please change the hyphen (-) into a minus sign (−, “U+2212”) in the figure, e.g., “-1” should be “−1”.
%\end{center}
\end{figure}

\begin{figure}[H] 
%\begin{center}
\includegraphics[width=0.45\linewidth,keepaspectratio]{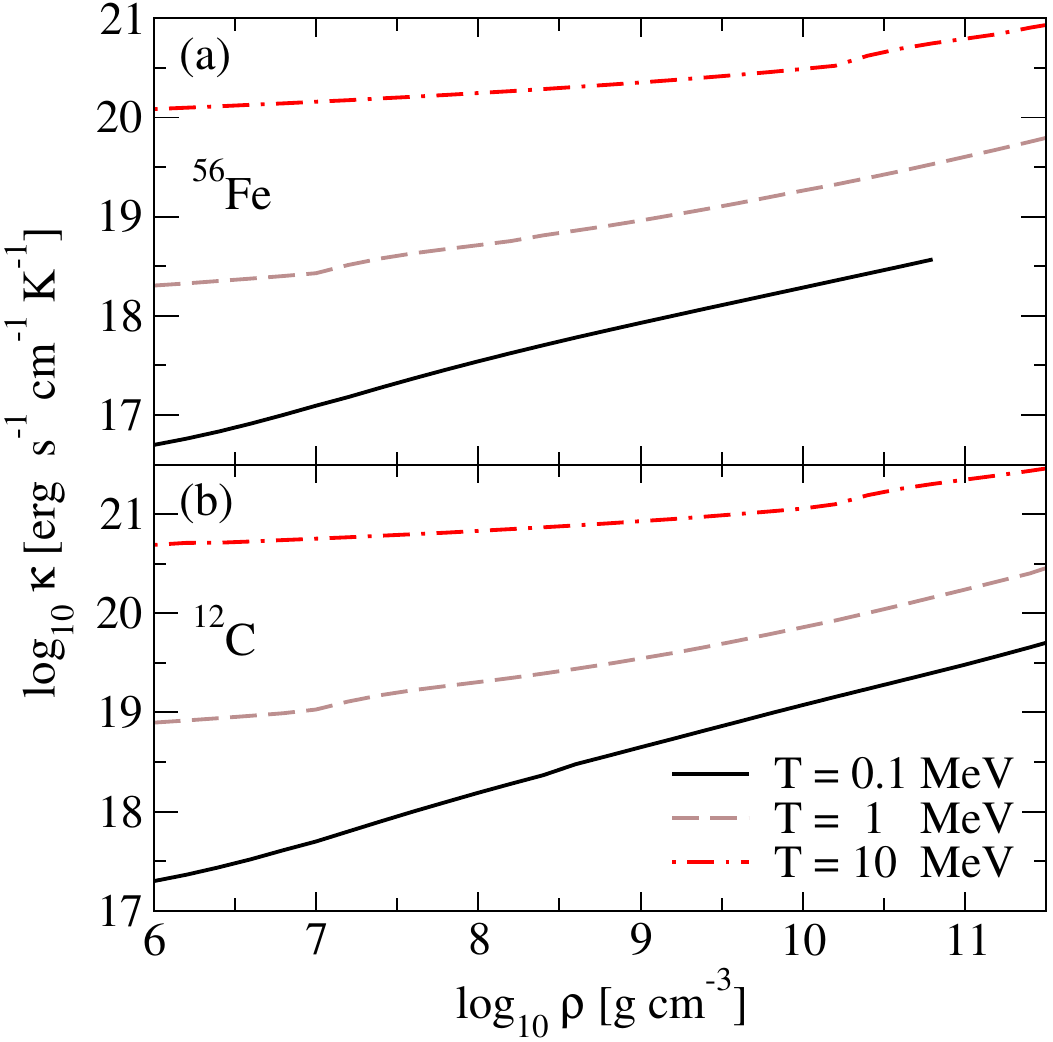}
\caption{{Dependence} of the longitudinal conductivity on density 
for various values of temperature for (\textbf{a}) $\isotope[56]{Fe}$; (\textbf{b}) $\isotope[12]{C}$. }%\MDPI: Please change the hyphen (-) into a minus sign (−, “U+2212”) in the figure, e.g., “-1” should be “−1”.  We moved this figure after it first citation in the whole text, please check and confirm
\label{fig:kappa_dens}
%\end{center}
\end{figure}

Figure~\ref{fig:tau_temp1} provides numerical values of the relaxation time as a function of temperature. As discussed above, $\tau$ depends on the electron energy; therefore it is evaluated at the Fermi energy in the degenerate regime 
(left panels) and at the thermal energy $\bar\varepsilon =3T$ in the nondegenerate (right panels) regime. 
Our results agree well with those of Nandkumar and Pethick~\cite{Nandkumar1984MNRAS} in the
degenerate regime, which are shown on our plots with circles.
{The values of their $I_1$ integral were read-off from the solid lines  shown in Figures \ref{fig:tau_temp1} and \ref{fig:kappa_dens}, which correspond to screening in the random-phase-approximation including vertex corrections. These were then used with Equation  (9) to compute the relaxation time. %Note that we use the bare electron mass in their formula.
} It is seen that $\tau$ decreases as a function of temperature in the degenerate regime and increases in the nondegenerate regime.
The decrease in the relaxation time with the temperature
in the degenerate regime is caused practically by the structure
factor $S(q)$. For comparison we evaluated also the relaxation time without taking into account the structure factor in Figure~\ref{fig:tau_temp1}. In the nondegenerate regime,
the temperature dependence of $\tau $ is dominated by the energy
increase in electrons with temperature, and the role of $S(q)$ is less
important, especially for light nuclei. This is due to the fact that when
$T\ge T_F$, \ie, electrons are nondegenerate, the ionic component
forms a weakly coupled Boltzmann gas (see Figure~\ref{fig:PhaseDiagram}).

  %------------------------------------------------------------------

%	\vspace{-3pt}

%----------------------------------------------------

\vspace{-10pt}
%----------------------------------------------------
%------------------------------------------------------------------

\subsection{Longitudinal Thermal Conductivity}
\label{sec:longitudinal_kappa}

We start the discussion of our numerical results with the density,
temperature, and composition dependence of the longitudinal thermal
conductivity $\kappa$.  Figure~\ref{fig:kappa_dens} shows the thermal
conductivity as a function of density for various temperatures for the
crustal matter consisting of $\isotope[56]{Fe}$ and $\isotope[12]{C}$ nuclei. The chosen values of
the temperature cover both the non-degenerate ($T=10$ MeV) and the
degenerate ($T=0.1$ MeV) regimes for electrons with the value $T=1$
MeV being representative for transition between these two regimes
(the transition density is around log$_{10}~\rho \simeq 7.5$ g
cm$^{-3}$ for both types of nuclei, see
Figure~\ref{fig:PhaseDiagram}). The thermal conductivity shows a
power-law increase in density $\kappa \propto \rho^{\alpha}$ with
$\alpha_{Fe} \simeq 0.37$ and $\alpha_{C}\simeq 0.43$
%$\alpha_{He} \simeq \alpha_{H}\simeq 0.45$ 
in the degenerate regime,
and $\alpha\simeq 0.08$ (for both nuclei) in the non-degenerate
regime. This behavior is similar to the density-dependence of the
electrical conductivity, which is a consequence of different density
and temperature dependence of the relaxation time in these regimes 
(see Ref.~\cite{Harutyunyan2016} for
details).

\textls[-10]{The main difference between the values of $\kappa$ for different
nuclei arise from the scaling
$\kappa\sim n_e \tau\sim Zn_i\tau\sim (Z/A)\tau\simeq 0.5\tau$,
$\tau^{-1}\sim Z^2n_i\sim Z^2 /A\simeq 0.5Z$, as seen from
Equation~\eqref{eq:relax_final},} therefore $\kappa\sim Z^{-1}$.  Thus, the
expected ratio of two conductivities is
%$\kappa_C/\kappa_{Fe}\simeq\tau_C/\tau_{Fe}\simeq Z_{Fe}/Z_C\simeq 4.3$, 
$\kappa_C/\kappa_{Fe}\simeq
4.3$ 
%$\kappa_{He}/\kappa_C\simeq 3$ and $\kappa_{H}/\kappa_{He}\simeq2$ 
(in the nondegenerate limit) , which are consistent with the results shown in Figure~\ref{fig:kappa_dens}.

\begin{figure}[H] 
%\begin{center}
\includegraphics[width=0.45\linewidth,keepaspectratio]{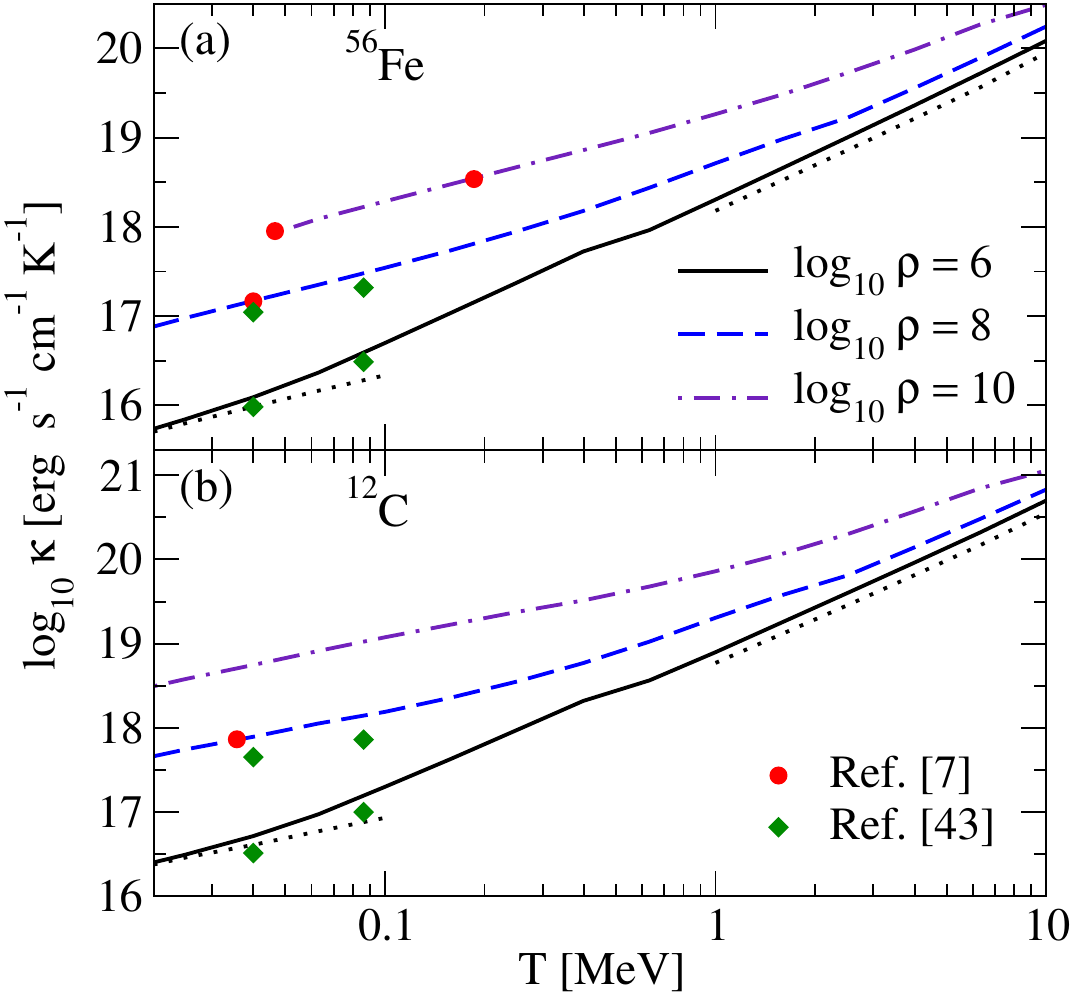}
\caption{{The} temperature dependence of the thermal conductivity 
 for three values of density for  $\isotope[56]  {Fe}$ (\textbf{a}) and  $\isotope[12]{C}$ (\textbf{b}). The
 dots represent the low- and high-temperature asymptotics. 
 {The red circles show the results of
Ref.~\cite{Nandkumar1984MNRAS} obtained from their values of $I_1$ integrals  given by solid lines in Figures \ref{fig:tau_temp1} and \ref{fig:kappa_dens} and Equation  (15). The green diamonds show the results listed in the table of Ref.~\cite{potekhin_code}.}}%\MDPI: Please change the hyphen (-) into a minus sign (−, “U+2212”) in the figure, e.g., “-1” should be “−1”. We moved this figure after it first citation in the whole text, please check and confirm
\label{fig:kappa_temp}
%\end{center}
\end{figure}

The temperature dependence of the thermal conductivity is shown in
Figure~\ref{fig:kappa_temp}. The dotted lines show the low- and
high-temperature asymptotics. The scaling of $\kappa$ with the
temperature is $\kappa\propto T^\gamma$ with
$\gamma_{Fe} \simeq 0.95$, $\gamma_{C} \simeq 0.8$
in the degenerate and
$\gamma\simeq 1.8$ in the non-degenerate regime.  In the degenerate
regime, the temperature dependence of $\kappa$ (or $\tau$) is stronger
for lighter elements, because at any given density and temperature,
the parameter $\Gamma$ is smaller for lighter elements
($\Gamma\sim Z^2/A^{1/3}$, \eg, $\Gamma_{Fe}/\Gamma_C\simeq 11$), 
and the structure factor $S(q)$
varies faster for small values of $\Gamma$, assuming that  
$a_i q$ is small.

%\vspace{-3pt}
%------------------------------------------------------

{For the sake of comparison, we show in Figure~\ref{fig:kappa_temp} 
the low-temperature results of Ref.~\cite{Nandkumar1984MNRAS}, which are in very good agreement with ours. Note that both computations take into account the electron--ion collisions but neglect the electron--electron ones. They also use fits to the same data for the static structure factor $S(q)$. However, they treat the screening effects encoded in the electron polarization tensor differently: Ref.~\cite{Nandkumar1984MNRAS} uses the highly degenerate limit of the random-phase-approximation of the polarization tensor including vertex corrections in the static limit; our study utilizes the hard-thermal-loop approximation to the dynamical polarization tensor (see Ref.~\cite{Harutyunyan2016} and references therein). We also show the results extracted from the table provided by Ref.~\cite{potekhin_code} for $T=8.64\cdot 10^8$~K and $T=10^9$~K, the largest value for temperature available. Our values of $\kappa$ are larger by about 30--40\% for $\isotope[56]{Fe}$ and  50--75\% for $\isotope[12]{C}$ 
compared to this work. The main difference is that Ref.~\cite{potekhin_code} includes electron--electron collisions in addition to electron--ion collisions. The structure factor used in this work differs from ours: it utilizes the fits of Ref.~\cite{Young1991} to hypernetted chain computations for one-component plasma, whereas we use data based on older Monte Carlo simulations of liquid structure function for the same system~\cite{Hansen1973,Galam1976,Tamashiro1999}. Similarly to the previous case, the treatments of the electron polarization tensor are not identical in terms of approximation and inclusion of dynamical effects.}

\subsection{Transverse and Hall Conductivities}

The transverse ($\kappa_0$) and Hall ($\kappa_1$) components of the
thermal conductivity depend on the value of the Hall parameter 
$\omega_c\tau$, which for homogeneous magnetic fields decreases with 
the density because of the decrease in relaxation time in any regime
as well as because of the decrease in $\omega_c $ in the degenerate regime. 
Thus, at low densities, the crust becomes more anisotropic, and for 
$\omega_c\tau\gg 1$ we have\vspace{-3pt}
%-----------------------------------------------------------------------
\bea\label{eq:kappa0_anisotrop}
\kappa_0\simeq \frac{\kappa}{(\omega_{c}\tau)^2}\simeq
%\frac{\pi^2 n_e\ep_F}{3 e^2\tau_F}\frac{T}{B^2}=
\left(\frac{\pi^2 n_eT}{3 eB}\right)^2\kappa^{-1}\ll\kappa.
\eea 
%-----------------------------------------------------------------------

\begin{figure}[H] 
%\begin{center}
\includegraphics[width=0.45\linewidth,keepaspectratio]{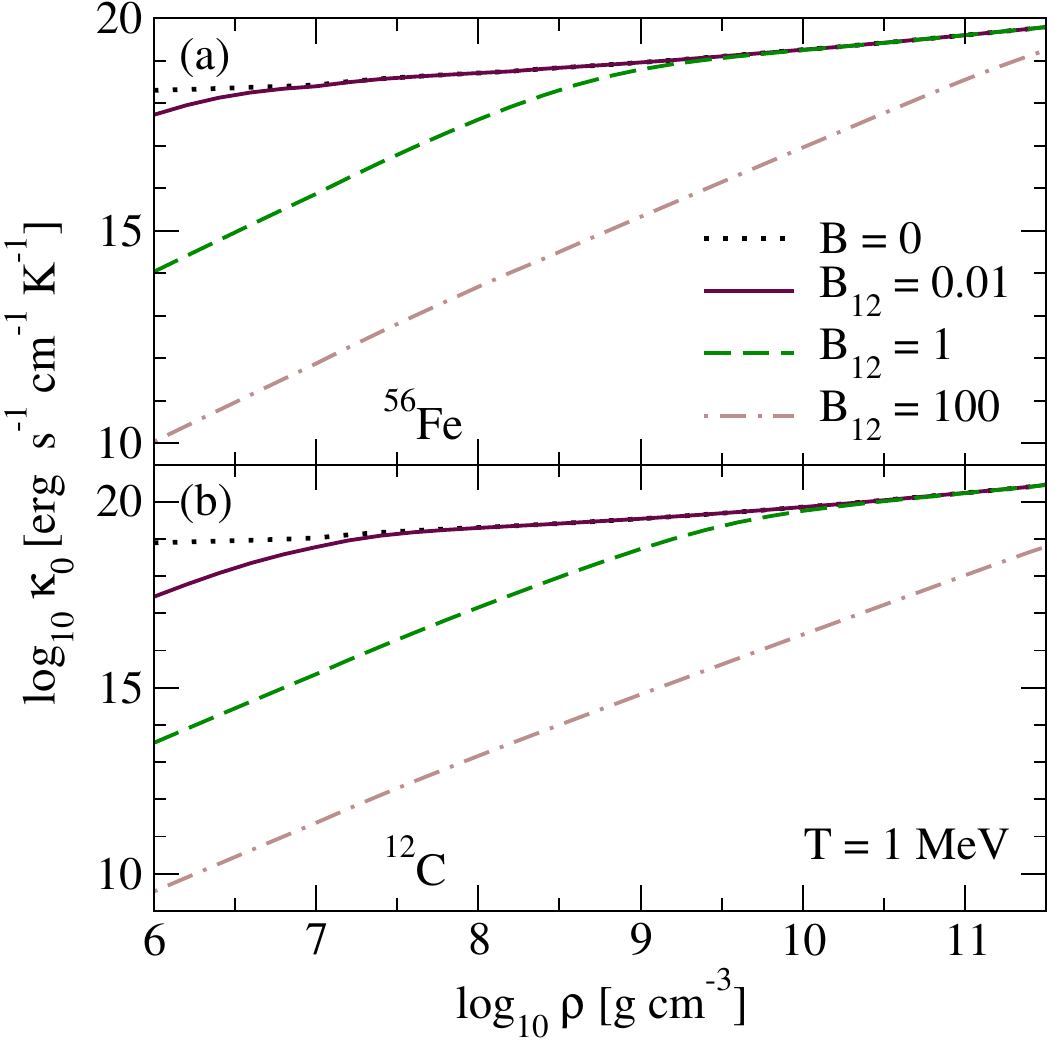}
\caption{{Dependence} of the transverse component of the thermal 
conductivity tensor on density for various values of the $B$-field 
for (\textbf{a}) $\isotope[56]{Fe}$; (\textbf{b}) $\isotope[12]{C}$. The temperature 
is fixed at $T=1$ MeV. The dotted lines show the limit of vanishing
magnetic field, \ie, the longitudinal conductivity. }%\MDPI: Please change the hyphen (-) into a minus sign (−, “U+2212”) in the figure, e.g., “-1” should be “−1”. We moved this figure after it first citation in the whole text, please check and confirm
\label{fig:kappa0_dens}
%\end{center}
\end{figure}

{Because} $\omega_c\tau$ decreases with the density, the component
$\kappa_0$ increases with density faster than $\kappa$ according to
scaling $\kappa_0\propto\rho^{\beta}$, $\beta\simeq 1.6$ in the
degenerate regime, and $\beta\simeq 1.85$ in the nondegenerate regime, and
tends to $\kappa_0$ in the high-density regime (see
Figure~\ref{fig:kappa0_dens}).  The scaling of $\kappa_0$ with respect to
$Z$ is the inverse of $\kappa$, i.e., $\kappa_0\sim\tau^{-1}\sim
Z$. Thus, $\kappa_0$ is smaller for lighter elements in the strongly
anisotropic regime. 
%The only exception is $\isotope[1]{H}$, for which the factor $n_e\propto Z/A$ is by a factor 2 larger than for the other elements. As a result, $\kappa_0$ is also by a factor larger for hydrogen compared to other elements (in particular helium) with $Z/A=1/2$.  
We see from Equation~\eqref{eq:kappa0_anisotrop} that for a
given density $\kappa_0/T\sim (\kappa/T)^{-1}$, therefore $\kappa_0/T$
shows an inverted temperature dependence at low densities.
%-------------------------------------------------------

%\vspace{-3pt}

%-------------------------------------------------------

For the Hall component, we find the following limiting expressions
%-----------------------------------------------------------
\bea\label{eq:kappa1_isotrop}
\kappa_1 &\simeq & (\omega_{c}\tau)\kappa
\simeq\frac{3e}{\pi^2}\frac{B}{n_eT}\kappa^2, 
\qquad\omega_c\tau\ll 1,\\
\label{eq:kappa1_anisotrop}
\kappa_1 &\simeq & \frac{\kappa}{\omega_{c}\tau}
\simeq\frac{\pi^2}{3e}\frac{n_e T}{B}, 
\qquad\omega_c\tau\gg 1.
\eea
%-----------------------------------------------------------

\begin{figure}[H] 
%\begin{center}
\includegraphics[width=0.45\linewidth,keepaspectratio]{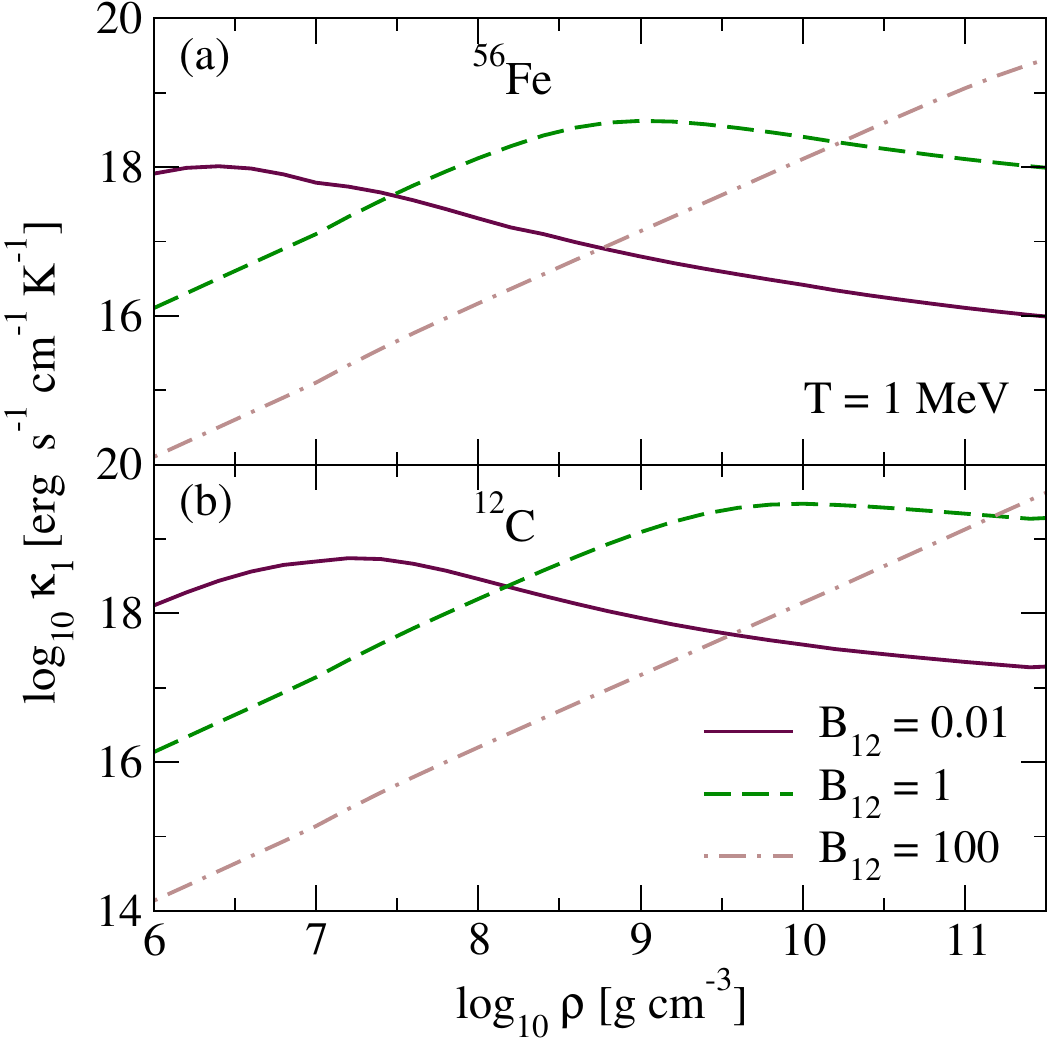}
\caption{{Dependence} of the Hall component of the thermal 
conductivity tensor on density for various values of the 
$B$-field for (\textbf{a}) $\isotope[56]{Fe}$; (\textbf{b}) $\isotope[12]{C}$. 
The temperature is fixed at $T=1$ MeV.  }%\MDPI: Please change the hyphen (-) into a minus sign (−, “U+2212”) in the figure, e.g., “-1” should be “−1”. We moved this figure after it first citation in the whole text, please check and confirm
\label{fig:kappa1_dens}
%\end{center}
\end{figure}

{At} low densities and/or large magnetic fields $\omega_c\tau\gg1$, and
$\kappa_1$ increases linearly with the density and temperature, but
does not depend on the type of nuclei. % as far as $A/Z\simeq 2$. It is by a factor of 2 larger for hydrogen with $A=Z=1$.
 At high densities,
$\kappa_1$ decreases slowly with the density, as implied by
Equation~\eqref{eq:kappa1_isotrop} (see Figure~\ref{fig:kappa1_dens}).  At the
point where $\omega_c\tau\simeq 1$, the transition between these two
regimes occurs, and $\kappa_1$ displays a maximum there.
%-------------------------------------------------------

%\vspace{-3pt}
%-------------------------------------------------------
In the isotropic region we have also $\kappa_1/T \propto (\kappa/T)^2$ and
$\kappa_1\propto Z^{-2}$ for $Z>1$, therefore $\kappa_1$ is larger for $\isotope[12]{C}$ by an order of
magnitude as compared to
$\isotope[56]{Fe}$. As $\tau$ is larger for light elements, the
anisotropic region for lighter elements is larger, and the maximum of
$\kappa_1$ is shifted to higher densities. Correspondingly, 
the maximum value of $\kappa$ is also larger for lighter elements.

\begin{figure}[H] 
%\begin{center}
\includegraphics[width=0.45\linewidth,keepaspectratio]{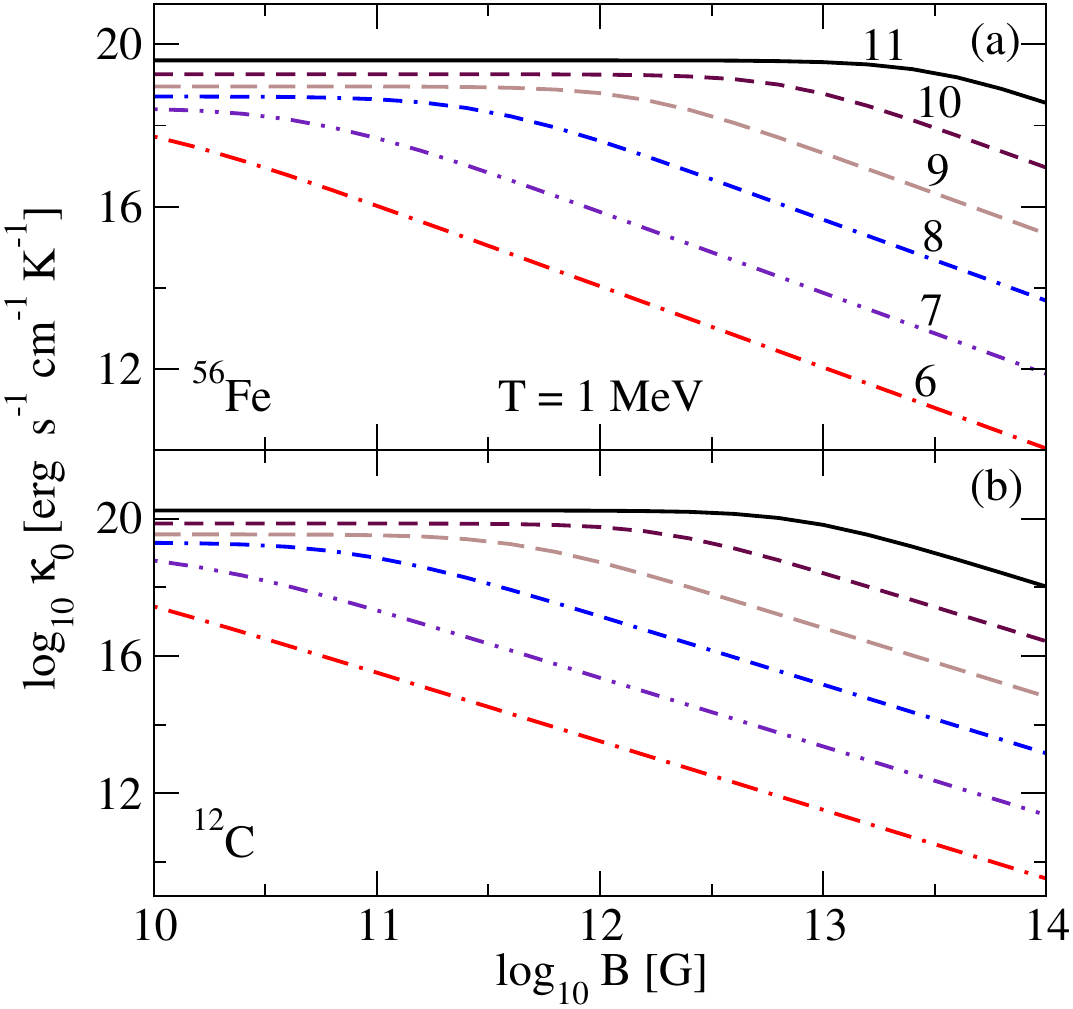}
\caption{{Dependence} of $\kappa_0$ on the magnetic field for (\textbf{a})
  $\isotope[56]{Fe}$; (\textbf{b}) $\isotope[12]{C}$ for various values of
  $\log_{10}\rho/[\textrm{g cm}^{-3}] $ as indicated in the plots. The
  temperature is fixed at $T=1$ MeV. }%\MDPI: Please change the hyphen (-) into a minus sign (−, “U+2212”) in the figure, e.g., “-1” should be “−1”. We moved this figure after it first citation in the whole text, please check and confirm
\label{fig:kappa0_b}
%\end{center}
\end{figure}

\begin{figure}[H] 
%\begin{center}
\includegraphics[width=0.45\linewidth,keepaspectratio]{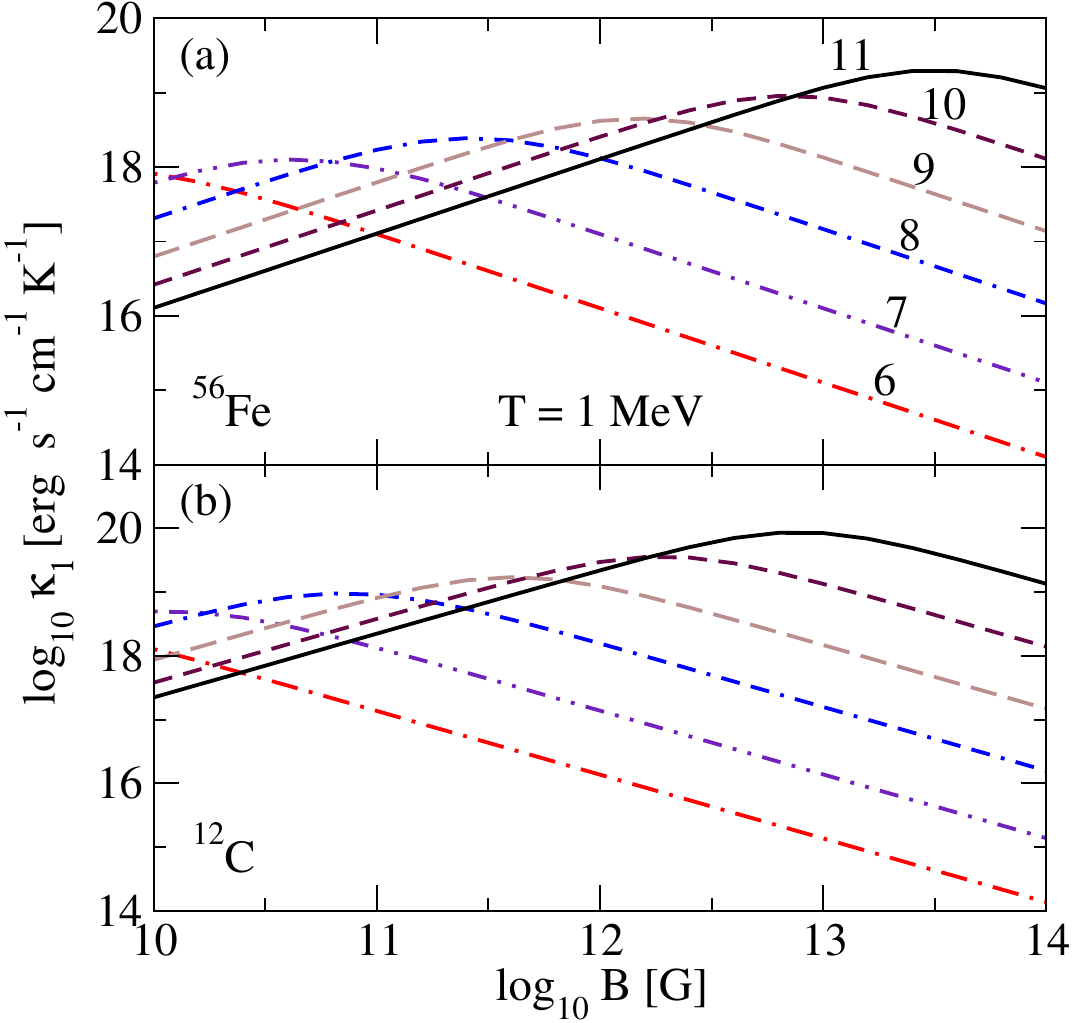}
\caption{{Dependence} of $\kappa_1$ on the magnetic field for (\textbf{}a)
  $\isotope[56]{Fe}$; (\textbf{b}) $\isotope[12]{C}$ for various values of
  $\log_{10}\rho/[\textrm{g cm}^{-3}] $ as indicated in the plots.
  The temperature is fixed at $T=1$ MeV. }%\MDPI: Please change the hyphen (-) into a minus sign (−, “U+2212”) in the figure, e.g., “-1” should be “−1”. We moved this figure after it first citation in the whole text, please check and confirm
\label{fig:kappa1_b}
%\end{center}
\end{figure}

Figures~\ref{fig:kappa0_b} and \ref{fig:kappa1_b} show the dependence
of $\kappa_0$ and $\kappa_1$ on the magnetic field.  For low magnetic
fields $\omega_c\tau\ll 1$, therefore $\kappa_0\simeq\kappa$ and it
does not depend on the value of the field. However, for
$\omega_c\tau\gtrsim 1$ $\kappa_0$ it starts decreasing as
$\kappa_0\propto B^{-2}$. The scaling of the Hall conductivity on the
magnetic field in the low-field and high-field limits is given by
$\kappa_1\propto B$ and $\kappa_1\sim B^{-1}$, respectively, and the
maximum is located around $\omega_c\tau\simeq 1$. The point of the
maximum shifts to lower magnetic fields with the decrease in density
and nucleus charge number.  For $B_{12}=1$, the crust is anisotropic at
densities $\rho\leq 10^9$ g cm$^{-3}$ for $\isotope[56]{Fe}$ and at $\rho\leq 10^{10}$ g cm$^{-3}$ for $\isotope[12]{C}$. 
 %and at $\rho\leq 3\cdot 10^{10}$ g cm$^{-3}$ for $\isotope[4]{He}$ and $\isotope[1]{H}$.  
 For magnetic fields $B_{12}\geq 10$, the outer crust
is almost entirely anisotropic for all nuclei.

%\vspace{-3pt}
%-------------------------------------------------------

\begin{figure}[H] 
%\begin{center}
\includegraphics[width=0.45\linewidth,keepaspectratio]{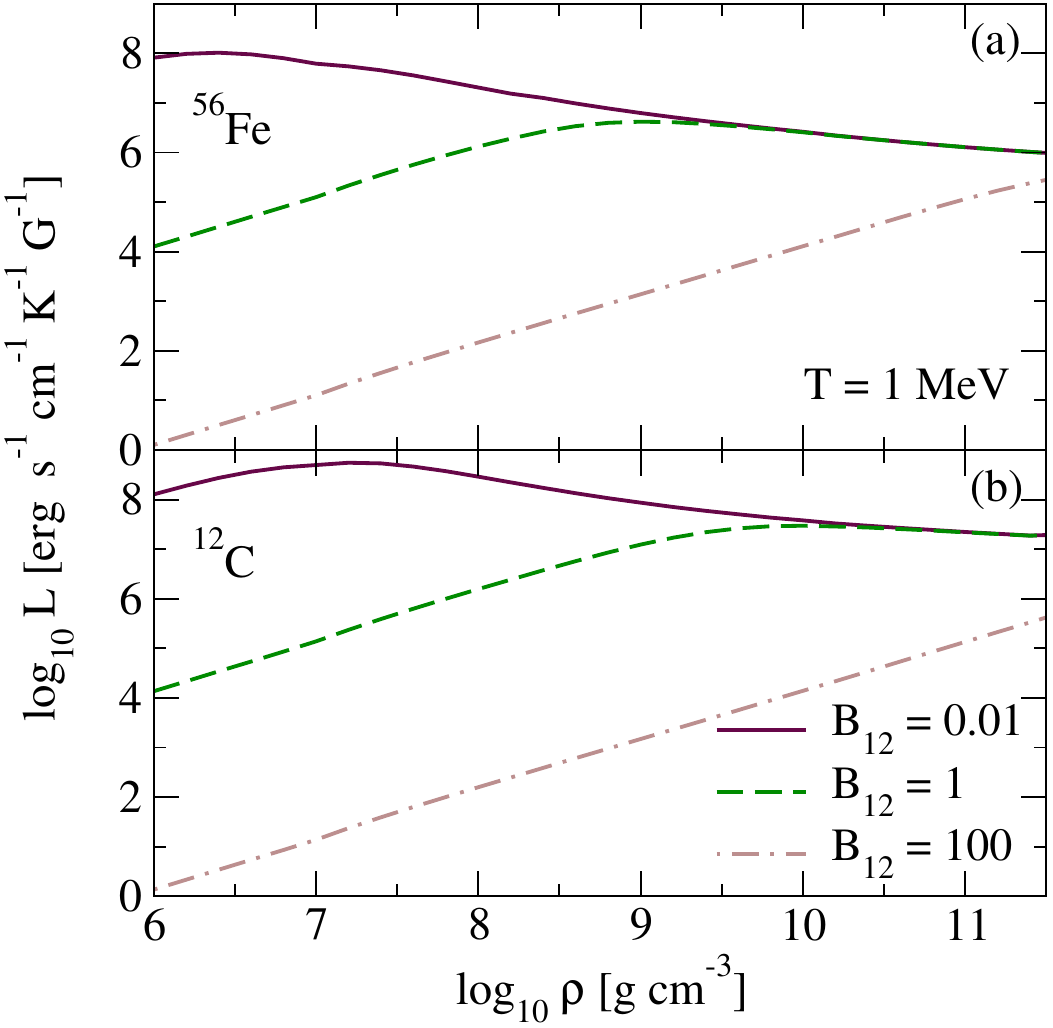}
\caption{{The} modulus of the 
Righi--Leduc coefficient as a function of density for various values of the 
$B$-field for (\textbf{a}) $\isotope[56]{Fe}$; (\textbf{b}) $\isotope[12]{C}$. 
The temperature is fixed at $T=1$ MeV. }%\MDPI: Please change the hyphen (-) into a minus sign (−, “U+2212”) in the figure, e.g., “-1” should be “−1”. We moved this figure after it first citation in the whole text, please check and confirm
\label{fig:Leduc_dens}
%\end{center}
\end{figure}

\begin{figure}[H] 
%\begin{center}
\includegraphics[width=0.45\linewidth,keepaspectratio]{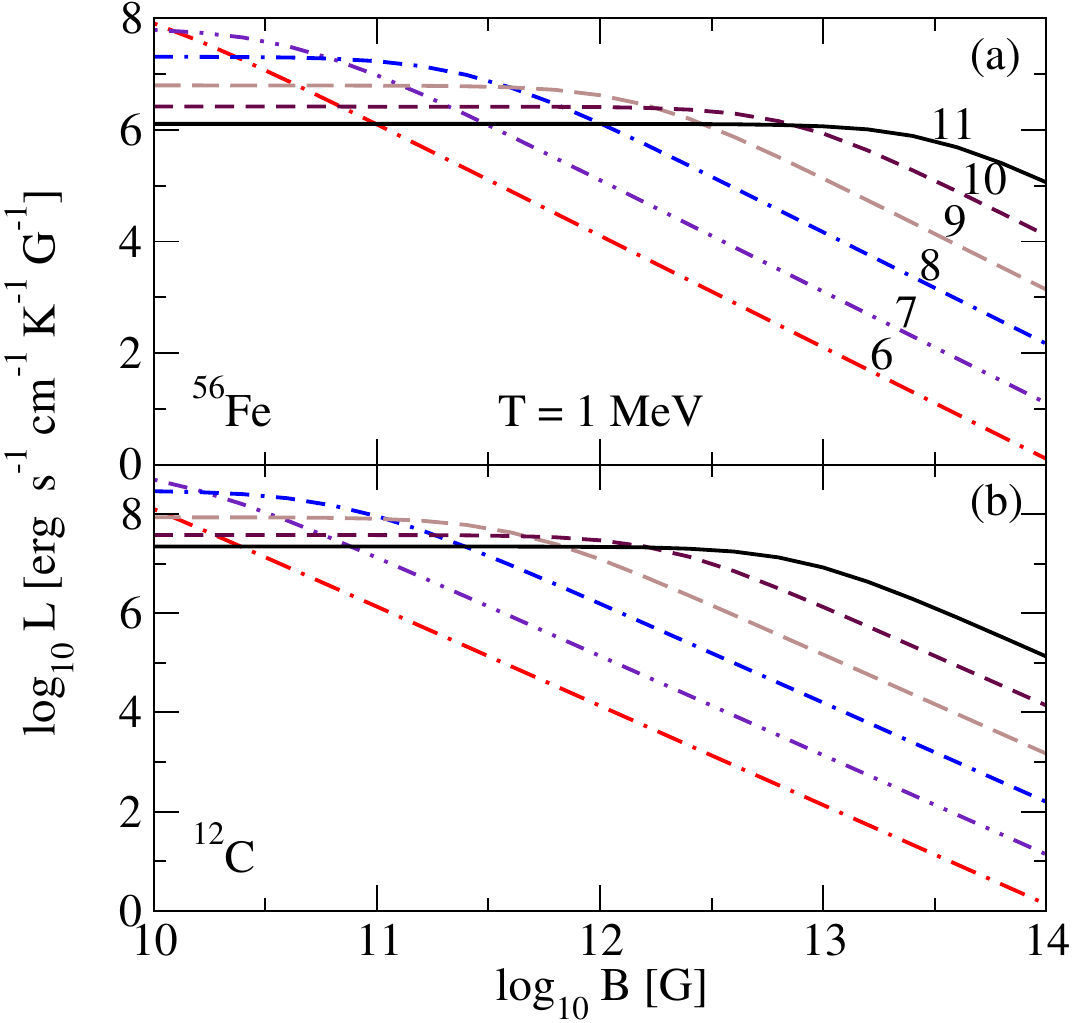}
\caption{\textls[-25]{{The} modulus of the Righi--Leduc coefficient as a function of magnetic field for (\textbf{a})
  $\isotope[56]{Fe}$; (\textbf{b}) $\isotope[12]{C}$ for various values of
  $\log_{10}\rho/[\textrm{g cm}^{-3}] $ as indicated in the plots.
  The temperature is fixed at $T=1$ MeV.} }%\MDPI: Please change the hyphen (-) into a minus sign (−, “U+2212”) in the figure, e.g., “-1” should be “−1”. We moved this figure after it first citation in the whole text, please check and confirm
\label{fig:Leduc_b}
%\end{center}
\end{figure}

%----------------Leduc-Righi coefficient 
Finally, we show the modulus of the Righi--Leduc coefficient $L=\kappa_1/B$ as defined after Equation~\eqref{eq:rho_comp} in Figures~\ref{fig:Leduc_dens} and \ref{fig:Leduc_b}, which 
has a negative sign for electrons.  This coefficient is responsible for the thermal Hall effect, which might affect the thermal evolution of the star if large thermal gradients are present perpendicular to the magnetic field lines and the Hall parameter is close to unity $\omega_c\tau\simeq 1$, where $L\simeq \kappa/2B$. From Equations~\eqref{eq:kappa1_isotrop} and \eqref{eq:kappa1_anisotrop}, we obtain for the limiting cases\vspace{-3pt}
%-----------------------------------------------------------
\bea\label{eq:L_isotrop}
L &\simeq & \frac{3e \kappa^2}{\pi^2 n_eT}, 
\qquad\omega_c\tau\ll 1,\\
\label{eq:L_anisotrop}
L &\simeq & \frac{\pi^2}{3e}\frac{n_e T}{B^2}, 
\qquad\omega_c\tau\gg 1.
\eea
%-------------------------------------------------------

\vspace{-12pt}
%-----------------------------------------------------------

\vspace{-6pt}
%-------------------------------------------------------

\vspace{-6pt}
%-----------------------------------------------------------

\subsection{Fits to the Conductivities}
\label{sec:fits}

We have performed fits to the (longitudinal) electrical and thermal conductivities using the formulae\vspace{-3pt}
%------------------------------------------------------------------
\begingroup
\makeatletter\def\f@size{8.6}\check@mathfonts
\def\maketag@@@#1{\hbox{\m@th\normalsize\normalfont#1}}%
\bea\label{eq:sigma_fit}
\sigma_{\rm fit} &=& \frac{1.5\times 10^{22}}{Z}
\left(
\frac{T_F}{1~{\rm MeV}}\right)^a\bigg(\frac{T}{T_F}\bigg)^{-b}
\bigg(\frac{T}{T_F}+d\bigg)^{c}~{\rm s}^{-1}.\\
\label{eq:kappa_fit}
\kappa_{\rm fit} &=& \frac{6\times 10^{19}}{Z}
\left(\frac{T}{1~{\rm MeV}}\right)
\left(\frac{\ep_F}{T_F}\right)^{0.16}\left(
\frac{T_F}{1~{\rm MeV}}\right)^{a'}\bigg(\frac{T}{T_F}\bigg)^{-b'}
\bigg(\frac{T}{T_F}+d'\bigg)^{c'}~{\rm erg~s^{-1}~cm^{-1}~K^{-1}}.
\eea
\endgroup

%----------------

{The} fit parameters in Formula \eqref{eq:sigma_fit} depend on the charge (proton number) of the nucleus via 
the following formula:\vspace{-3pt}
%----------------
\bea\label{eq:fit_constants_sigma}
a(Z) &=& 0.924 - 0.009 \log Z + 0.003 \log^2 Z,\\
b(Z) &=& 0.507 - 0.028 \log Z - 0.025 \log^2 Z,\\
c(Z) &=& 1.295 - 0.018 \log Z - 0.022 \log^2 Z,\\
d(Z) &=& 0.279 + 0.056 \log Z + 0.035 \log^2 Z.
\eea
%----------------

{The} fit parameters in Formula \eqref{eq:kappa_fit} are given by \vspace{-3pt}
%----------------
\bea\label{eq:fit_constants_kappa}
a'(Z) &=& 0.974 + 0.002 \log Z + 0.003 \log^2 Z,\\
b'(Z) &=& 0.393 - 0.009 \log Z - 0.030 \log^2 Z,\\
c'(Z) &=& 1.112 + 0.001 \log Z - 0.025 \log^2 Z,\\
d'(Z) &=& 0.224 + 0.028 \log Z + 0.042 \log^2 Z.
\eea
%--------------------------------------------------

{The} relative error of the fit Formulas \eqref{eq:sigma_fit} and \eqref{eq:kappa_fit}
%defined as $\gamma=100\vert\sigma^{\rm fit}-\sigma\vert/\sigma$ 
is 10\% in the temperature range 0.4~MeV $\le T\le$ 10 MeV. 
The density dependence in these formulas is by means of the electron Fermi temperature, which in general has the form $T_F=0.511\big[\sqrt{1+(Z\rho_6/A)^{2/3}}-1\big]$ MeV with $\rho_6 = \rho/(10^6\,{\rm g\  cm}^{-3})$. For
ultrarelativistic electrons, this simplifies to
$T_F=0.511 (Z\rho_6/A)^{1/3}\simeq 0.4\rho_6^{1/3}$ MeV, where the second relation is valid for $Z/A \simeq 0.5$.
Substituting this into Equations~\eqref{eq:sigma_fit} and \eqref{eq:kappa_fit}, we obtain fit
formulas with explicit dependence on density
(ultrarelativistic electrons)
%-----------------------------------------------------------
\bea\label{eq:sigma_fit_ultra}
\sigma_{\rm fit} & \simeq & C Z^{-1}
 \rho_6^{(a+b)/3}T_1^{-b} 
\left(T_1\rho_6^{-1/3}+yd\right)^{c}~{\rm s}^{-1},\\
\label{eq:kappa_fit_ultra}
\kappa_{\rm fit} & \simeq & C' Z^{-1}
 \rho_6^{(a'+b')/3}T_1^{1-b'} 
\left(T_1\rho_6^{-1/3}+yd'\right)^{c'}~{\rm erg~s^{-1}~cm^{-1}~K^{-1}},
\eea
%------------------------------------------------------------------
with $T_1\equiv T/(1~{\rm MeV})$,
$C =1.5\times 10^{22}\times y^{a+b-c}$, $C' =6\times 10^{19}\times y^{a+b-c}$, and $y = 0.4$.% for $Z\ge 2$; $y = 0.5$ for $Z=1$.

The average relaxation time $\tau$ and the Hall parameter $x\equiv \omega_c\tau$ can be fitted by the approximate formula\vspace{-3pt}
%------------------------------------------------------------------
\bea\label{eq:tau_fit}
\tau_{\rm fit} &=& 1.94\times 10^{-16}\times
\frac{A}{Z^2\rho_6}\left(
\frac{T_F}{1~{\rm MeV}}\right)^a\bigg(\frac{T}{T_F}\bigg)^{-b}
\bigg(\frac{T}{T_F}+d\bigg)^{c}\nonumber\\
&& \times
\left[\ep_{F1}\left(1+\frac{T}{T_F}\right)^{-1} +3T_1\left(1+\frac{T_F}{T}\right)^{-1}\right]~{\rm s},\\
\label{eq:Hall_fit}
x &=& 1.74\times 10^{3}\times\frac{A}{Z^2}\frac{B_{12}}{\rho_6}\left(
\frac{T_F}{1~{\rm MeV}}\right)^a\bigg(\frac{T}{T_F}\bigg)^{-b}
\bigg(\frac{T}{T_F}+d\bigg)^{c},
\eea
%-----------------------------------------------------------
where $\ep_{F1}\equiv \ep_F/(1~{\rm MeV})$.

For the transverse and Hall components of conductivities, the following formulas can be used\vspace{-3pt}
%------------------------------------------------------------------
\bea\label{eq:sigma01_fit}
\sigma_0^{\rm fit}=\frac{\nu\sigma_{\rm fit}}{1+(\nu x)^2},\qquad
\sigma_1^{\rm fit}=\frac{x\sigma_{\rm fit}}{1+x^2},\\
\label{eq:kappa01_fit}
\kappa_0^{\rm fit}=\frac{\nu\kappa_{\rm fit}}{1+(\nu x)^2},\qquad
\kappa_1^{\rm fit}=\frac{x\kappa_{\rm fit}}{1+x^2},
\eea
%------------------------------------------------------------------
where $\nu = ({T_F}/{\varepsilon_F})^{0.16}$.
 The relative error of Formulas~\eqref{eq:sigma01_fit} and 
\eqref{eq:kappa01_fit} is 15\% and \linebreak20\%,~respectively.

\section{Conclusions}
\label{sec:conclusions}

Motivated by the potential importance of the transport phenomena in
modeling the behaviour of finite-temperature electron-ion plasma in
hot neutron stars, white dwarfs, and binary neutron star mergers, we
have calculated the thermal conductivity and thermal Hall effect in
such plasma at densities corresponding to the outer crusts of neutron
stars and the interiors of white dwarfs, at temperatures from the melting
temperature of the solid $\sim$$10^9$~K up to $10^{11}$~K (10 MeV). This
temperature range is relevant to the liquid state of electron-ion
plasma and covers the transition from degenerate to non-degenerate
regimes for electrons. In this regime, electron transport dominates,
with electrons scattering off correlated nuclei via screened
electromagnetic forces. The plasma correlations in the liquid state
are accounted for using the ion structure-function derived from Monte
Carlo simulation data of one-component plasma. We have incorporated
hard-thermal-loop quantum electrodynamics polarization
susceptibilities in the low-frequency limit, combined with a
nonzero-temperature Debye screening length, which serves as a good
approximation in scenarios where inelastic processes are suppressed by
the large mass of nuclei. The Boltzmann kinetic equation is solved
using the relaxation time approximation.

Our  study of  the thermal conductivity of crustal matter focused on 
$\isotope[56]{Fe}$ and $\isotope[12]{C}$ nuclei. The conductivity 
shows a power-law increase in density $\kappa \propto \rho^{\alpha}$ 
in both regimes of
degenerate (with $\alpha \simeq 0.4$) and non-degenerate electrons
(with $\alpha\simeq 0.08$), similar to the density-dependence of the
electrical conductivity studied in Ref.~\cite{Harutyunyan2016}.

The scaling of $\kappa$ with the temperature is
$\kappa\propto T^\gamma$ with $\gamma\simeq 0.9$ in the degenerate and
$\gamma\simeq 1.8$ in the non-degenerate regime. We show also that
the behavior of the ratio $\kappa/T$ is rather similar to that of the
electrical conductivity. Specifically, $\kappa/T$ has a minimum which
lies in the range of  temperatures  $ 0.1\,T_F\le  T\le  0.15\,T_F$.
The occurrence of the minimum is observed across the entire density range we considered. This phenomenon is linked to the expansion of phase space caused by rising temperature, which is effective at any given fixed density. {Our results show good agreement with those of \cite{Nandkumar1984MNRAS} for the relaxation time and the thermal conductivity in the low-temperature regime, where a comparison is possible (see Figures~\ref{fig:tau_temp1} and \ref{fig:kappa_temp}). 
There are, however, some differences in the input physics, notably the use of different approximation for the electron polarization tensor in Ref.~\cite{Nandkumar1984MNRAS}. Additionally, we compared our thermal conductivity calculations with the tabulated results in \cite{potekhin_code}.  As explained already in Section~\ref{sec:longitudinal_kappa}, we find larger values of $\kappa$ by about 30--40\% for $\isotope[56]{Fe}$ and   50--75\% for $\isotope[12]{C}$ nuclei (see Figure~\ref{fig:kappa_dens}), mainly because our neglect of the electron--electron collisions. Additionally, both the static structure functions for one-component plasma and the electron polarization tensors have different origins (see Section~\ref{sec:longitudinal_kappa}), which might account for any potential differences.}

We studied also the transverse ($\kappa_0$) and the Hall ($\kappa_1$)
components of the thermal conductivity,
as well as the Righi--Leduc coefficient $L=\kappa_1/B$ 
for nonquantizing magnetic
fields $B\le 10^{14}$~G. These components depend on the value of the
Hall parameter $\omega_c\tau$, which decreases with density. As a
result, the low-density regions of the crust are more anisotropic,
meaning that the thermal conduction is suppressed in the directions
transverse to the magnetic field.  This anisotropy sets in at the
density where $\omega_c\tau\simeq 1$. In the high-field limit, we find
the following scalings of the transverse $\kappa_0\propto B^{-2}$ and
Hall $\kappa_1\sim B^{-1}$ conductivities on the magnetic field $B$.
{Our numerical results can be implemented in hydrodynamics simulations of warm compact stars also by means of accurate fit formulas. }
Large-scale simulations of the early phases of hot proto-neutron stars and post-merger remnants can shed light on the role of thermal conductivity and thermal Hall effect on the turbulent fluid dynamics of these objects under the combined influence of strong magnetic fields and thermal gradients, potentially uncovering complex feedback loops between various factors.

While we focused here on thermal conductivity and the thermal Hall effect,
we anticipate that the thermopower could play an equally important role
in the dissipative dynamics of astrophysical electron-ion plasma in
strong fields. We relegate the study of the thermopower as well as of
the astrophysical implications of the present study to a \mbox{future publication.}

\vspace{6pt}

\authorcontributions{{The authors contributed equally to this work.  All authors have read and agreed to the published version of the manuscript. } %MDPI: For research articles with several authors, a short paragraph specifying their individual contributions must be provided. The following statements should be used ``Conceptualization, X.X. and Y.Y.; methodology, X.X.; software, X.X.; validation, X.X., Y.Y. and Z.Z.; formal analysis, X.X.; investigation, X.X.; resources, X.X.; data curation, X.X.; writing---original draft preparation, X.X.; writing---review and editing, X.X.; visualization, X.X.; supervision, X.X.; project administration, X.X.; funding acquisition, Y.Y. All authors have read and agreed to the published version of the manuscript.'', please turn to the \href{http://img.mdpi.org/data/contributor-role-instruction.pdf}{CRediT taxonomy} for the term explanation. Authorship must be limited to those who have contributed substantially to the work~reported.
}

\funding{{This research was funded by Deutsche
    Forschungsgemeinschaft grant number SE 1836/5-3, Polish National
    Science Centre (NCN) grant 2020/37/B/ST9/01937, Higher Education and Science Committee (HESC) of the Republic of Armenia Grant No. 24RL-1C010.} %MDPI: Please add: ``This research received no external funding'' or ``This research was funded by NAME OF FUNDER grant number XXX.'' and and ``The APC was funded by XXX''. Check carefully that the details given are accurate and use the standard spelling of funding agency names at \url{https://search.crossref.org/funding}, any errors may affect your future funding.
}

\dataavailability{{The data underlying this article will be shared at a reasonable request by the corresponding author.} %MDPI: We encourage all authors of articles published in MDPI journals to share their research data. In this section, please provide details regarding where data supporting reported results can be found, including links to publicly archived datasets analyzed or generated during the study. Where no new data were created, or where data is unavailable due to privacy or ethical restrictions, a statement is still required. Suggested Data Availability Statements are available in section ``MDPI Research Data Policies'' at \url{https://www.mdpi.com/ethics}.
} 

\acknowledgments{A.H. acknowledges the hospitality of Frankfurt Institute for
Advanced Studies at Goethe University where part of this work was
carried out. }%\MDPI: To AE: please confirm if the funding information in the Acknowledgments Section should be moved to the Funding Section.

\conflictsofinterest{{The authors declare no conflicts of interest. } %MDPI: Declare conflicts of interest or state ``The authors declare no conflict of interest.'' Authors must identify and declare any personal circumstances or interest that may be perceived as inappropriately influencing the representation or interpretation of reported research results. Any role of the funders in the design of the study; in the collection, analyses or interpretation of data; in the writing of the manuscript; or in the decision to publish the results must be declared in this section. If there is no role, please state ``The funders had no role in the design of the study; in the collection, analyses, or interpretation of data; in the writing of the manuscript; or in the decision to publish the results''.
}

\begin{adjustwidth}{-\extralength}{0cm}

\reftitle{References}

%%%%%%%%%%%%%%%%%%%%%%%%%%%%%%%%%%%%%%%%%%
\PublishersNote{}
\end{adjustwidth}
\end{document}